\documentclass[10pt,twocolumn]{IEEEtran}
\usepackage{indentfirst}
\usepackage[dvips]{graphicx}
\usepackage{amsfonts}
\usepackage{multirow}
\usepackage{amsmath,amsthm,amssymb}
\usepackage{color,changepage}
\usepackage{algorithm}
\usepackage{algorithmic}
\usepackage{bbm}
\usepackage{verbatim}
\usepackage{makecell}
\usepackage{subfigure}
\usepackage{mathrsfs}
\usepackage{arydshln}
\usepackage{cases}
\usepackage{threeparttable}
\usepackage{extarrows}
\usepackage{array}
\usepackage{bm}
\usepackage{pifont}
\usepackage[T1]{fontenc}
\usepackage{threeparttable}
\usepackage{booktabs}
\usepackage{mdwlist}
\usepackage{hyperref}
\usepackage{cite}
\allowdisplaybreaks[4]
\usepackage[dvipsnames]{xcolor}
\usepackage[most]{tcolorbox}

\definecolor{newcolor}{rgb}{0.5,0,1}

\newcolumntype{I}{!{\vrule width 2pt}}

\newtheorem{Theorem}{Theorem}
\newtheorem{Corollary}{Corollary}
\newtheorem{Definition}{Definition}

\newtheorem{Lemma}{Lemma}

\theoremstyle{remark}
\newtheorem{Remark}{Remark}
\newtheorem{Example}{Example}

\DeclareMathAlphabet{\mathpzc}{OT1}{pzc}{m}{it}

\definecolor{newcolor}{rgb}{0.5,0,1}
\newcommand{\minitab}[2][l]{\begin{tabular}{#1}#2\end{tabular}}

\begin{document}
\title{
A Systematic Approach towards Efficient Private Matrix Multiplication
}
\author{Jinbao Zhu and Songze Li
\thanks{
This work was  supported in part by the National Nature Science Foundation of China (NSFC) under Grant 62106057.

Jinbao Zhu is with the Thrust of Internet of Things, The Hong Kong University of Science and Technology (Guangzhou), Guangzhou 510006, China (e-mail: jbzhu@ust.hk).

Songze Li is with the Thrust of Internet of Things, The Hong Kong University of Science and Technology (Guangzhou), Guangzhou 510006, China, and also with the Department of Computer Science and Engineering, The Hong Kong University of Science and Technology, Hong Kong SAR, China (e-mail: songzeli@ust.hk).
}
}

\maketitle

\begin{abstract}
We consider the problems of Private and Secure Matrix Multiplication (PSMM) and Fully Private Matrix Multiplication (FPMM), for which matrices privately selected by a master node are multiplied at distributed worker nodes without revealing the indices of the selected matrices, even when a certain number of workers collude with each other. We propose a novel systematic approach to solve PSMM and FPMM with colluding workers, which leverages solutions to a related Secure Matrix Multiplication (SMM) problem where the data (rather than the indices) of the multiplied matrices are kept private from colluding workers. Specifically, given an SMM strategy based on polynomial codes or Lagrange codes, one can exploit the special structure inspired by the matrix encoding function to design private coded queries for PSMM/FPMM, such that the algebraic structure of the computation result at each worker resembles that of the underlying SMM strategy. Adopting this systematic approach provides novel insights in private query designs for private matrix multiplication, substantially simplifying the processes of designing PSMM and FPMM strategies. Furthermore, the PSMM and FPMM strategies constructed following the proposed approach outperform the state-of-the-art strategies in one or more performance metrics including recovery threshold (minimal number of workers the master needs to wait for before correctly recovering the multiplication result), communication cost, and computation complexity, demonstrating a more flexible tradeoff in optimizing system efficiency.
\end{abstract}


\section{Introduction}\label{Introduction}
\IEEEPARstart{I}{n} the era of Big Data, performing computationally intensive tasks on a single machine is becoming infeasible due to limited processing power and storage space. As an efficient solution, distributed computing has emerged as a natural approach to overcome such limitations, by partitioning the large computing task into smaller sub-tasks, and outsourcing them to many distributed worker nodes. However, scaling out the computation across distributed workers is also faced with efficiency challenges including additional communication overhead compared to centralized processing, and prolonged task execution time due to slow or delay-prone worker nodes, known as the \textit{straggler effect}~\cite{Tail1,Tail3}. Meanwhile, distributing sensitive raw data across worker nodes may raise serious security and privacy concerns. Therefore, designing computation and communication efficient strategies that are robust to straggler effect, while providing data privacy and security is of vital importance for distributed computing applications.

\begin{figure}[htbp]
\centering
    \includegraphics[width=8.5cm]{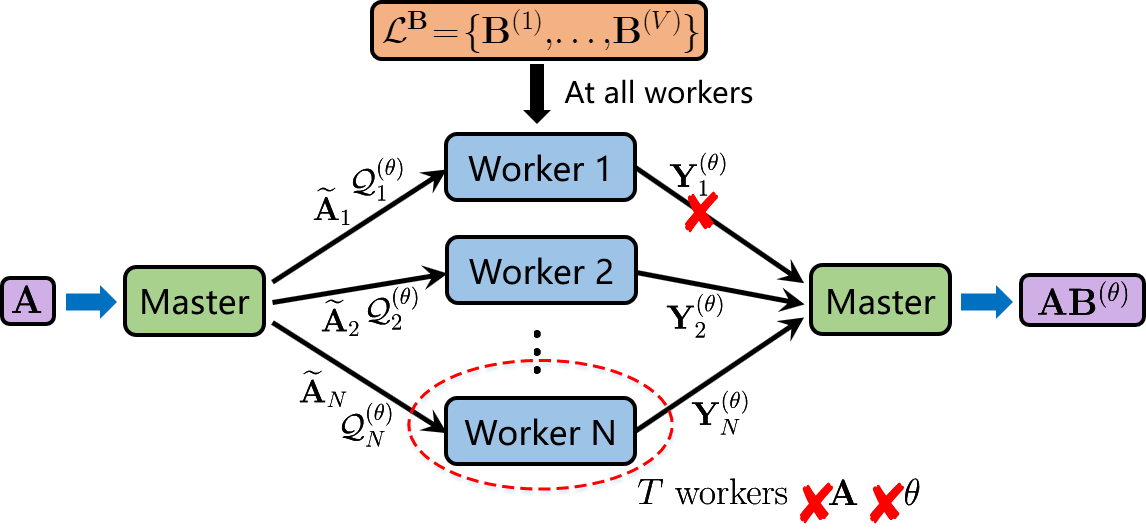}
    \vspace{-2mm}
    \caption{System model for private and secure matrix multiplication. The master sends a coded matrix $\widetilde{{\bf A}}_i$ and a query ${\cal Q}_i^{(\theta)}$ to worker $i$ with the private index $\theta$. Each worker $i$ uses $\widetilde{{\bf A}}_i$, ${\cal Q}_i^{(\theta)}$, and the public library ${\cal L}^{{\bf B}}$ to compute a response ${\bf Y}_i^{(\theta)}$. The master must be able to decode the product $\mathbf{A}\mathbf{B}^{(\theta)}$ from the responses of servers in the presence of stragglers.}
    \label{PSMM}
\end{figure}

Matrix multiplication, as one of the key building blocks in various engineering applications like machine learning and big data analysis, is typically carried out in a distributed manner for practically sized input matrices~\cite{ABFT,herault2015fault,CIT-103,lee2017high}. In this paper we focus on improving the computation and communication efficiency of two distributed private matrix multiplication problems, over a distributed computing system consisting of a master node and $N$ worker nodes. For the first \emph{Private and Secure Matrix Multiplication} (PSMM) problem, as illustrated in Fig. \ref{PSMM}, the master owns a confidential matrix $\mathbf{A}$ and all workers have access to a library $\mathcal{L}^{\mathbf{B}}$ of $V$ public matrices $\mathbf{B}^{(1)},\mathbf{B}^{(2)},\ldots,\mathbf{B}^{(V)}$. The goal of the master is to compute the product $\mathbf{A}\mathbf{B}^{(\theta)}$ for some $\theta\in\{1,2,\ldots,V\}$ from the distributed system, while keeping the index $\theta$ private and the matrix $\mathbf{A}$ secure from any up to $T$ colluding workers. 

\begin{figure}[ht]
    \centering
    \includegraphics[width=8.3cm]{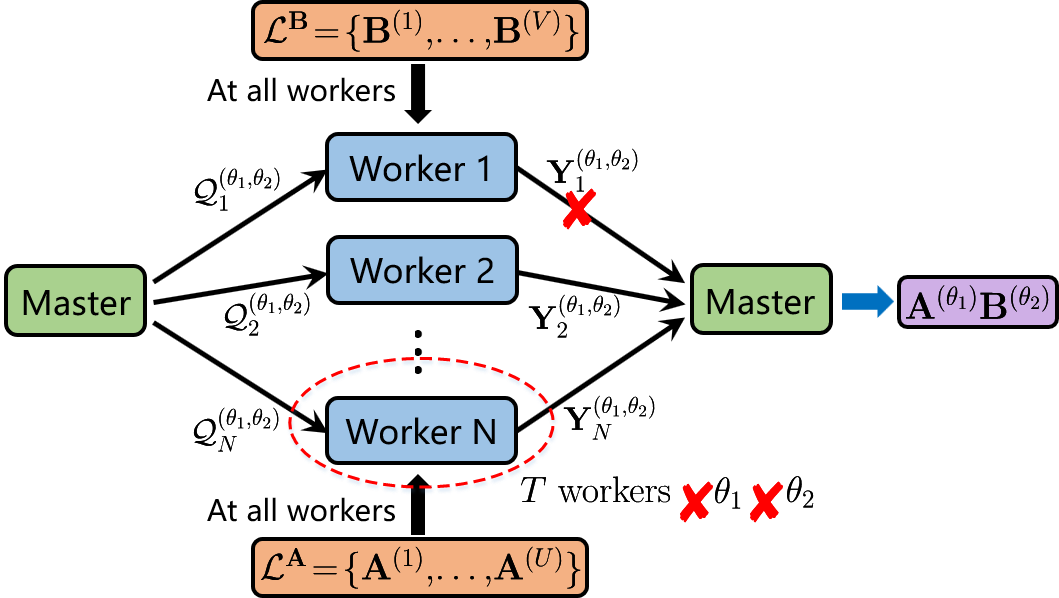}
    \vspace{-2mm}
    \caption{System model for fully private matrix multiplication. The master sends a query ${\cal Q}_i^{(\theta_1, \theta_2)}$ to worker $i$ with the private indices $\theta_1$ and $\theta_2$. Each worker $i$ uses ${\cal Q}_i^{(\theta_1, \theta_2)}$, and the public libraries ${\cal L}^{{\bf A}}$ and ${\cal L}^{{\bf B}}$ to compute a response ${\bf Y}_i^{(\theta)}$. The master must be able to decode the product $\mathbf{A}^{(\theta_1)}\mathbf{B}^{(\theta_2)}$ from the responses of servers in the presence of stragglers. }
    \label{FPMM}
\end{figure}

To do that, the master sends an encoded version of the matrix $\mathbf{A}$ to each worker, along with a query that instructs the worker to encode the library $\mathcal{L}^{\mathbf{B}}$ and compute a response for the master. To mitigate the influence of stragglers, the master only waits for the responses from a subset of fastest workers to recover the desired product $\mathbf{A}\mathbf{B}^{(\theta)}$, where the minimum number of successful computing workers that the master needs to wait for is referred to as \emph{recovery threshold}. In the second problem of interest shown in Fig. \ref{FPMM}, referred to as \emph{Fully Private Matrix Multiplication} (FPMM), the matrix $\mathbf{A}$ is selected from another public library $\mathcal{L}^{\mathbf{A}}$ of $U$ matrices $\mathbf{A}^{(1)},\mathbf{A}^{(2)},\ldots,\mathbf{A}^{(U)}$ that are shared by all workers along with the library $\mathcal{L}^{\mathbf{B}}$. In this case, the master wishes to compute $\mathbf{A}^{(\theta_1)}\mathbf{B}^{(\theta_2)}$ for some $\theta_1\in\{1,2,\ldots,U\}$ and $\theta_2\in\{1,2,\ldots,V\}$, without revealing any information about the indices $\theta_1$ and $\theta_2$ to any $T$ colluding workers. Private matrix multiplication~\cite{PSMM:2,PSMM:1} has a wide range of application scenarios in practice. For instance, consider a recommender system based on collaborative filtering, where recommendations are generated by computing the product of two matrices, one describing the profiles of the users, and another one representing the profiles of the items. Given that the user profile matrix may reveal the users' private information, and the queried user and item indices may leak the privacy of the recommendation requester, data security and query  privacy should be provided by the recommendation service.

Secure Matrix Multiplication (SMM) is another problem that is related to the interested PSMM and FPMM problems. In SMM, the master wishes to compute the product of two owned matrices $\mathbf{A}$ and $\mathbf{B}$ in the distributed system, without revealing anything about $\mathbf{A}$ and $\mathbf{B}$ to the workers. Computing strategies for the SMM problem, based on how the matrices ${\bf A}$ and ${\bf B}$ are securely encoded, can be categorised into SMM based on polynomial codes \cite{PSMM:2,Rouayheb_secure_code,Tandon_secure_code,EP_SMC,Kakar_secure_code,Kakar_and_Khristoforov,Gunduz20,Zhu_SDMM} and SMM based on Lagrange codes \cite{Qian_Yu,batch_matrix}, where polynomial codes \cite{Polynomial_code,MatDot_code,EP_code,GpolyDot} and Lagrange codes \cite{LCC} are constructed by leveraging the algebraic structure of polynomial functions and Lagrange interpolate polynomials, respectively.
An essential component behind these coded strategies is to construct appropriate encoding functions of $\mathbf{A}$ and $\mathbf{B}$, such that the desired product $\mathbf{A}\mathbf{B}$ can be recovered by interpolating a polynomial from worker responses.
The state-of-the-art strategies for SMM based on polynomial codes and Lagrange codes are reflected in \cite{Zhu_SDMM,EP_SMC} and \cite{Qian_Yu}, respectively. Having observed the similarities between PSMM/FPMM and SMM in requiring privacy-preserving matrix multiplication, and their key difference that whether or not queries for the workers are needed, we are interested in the question:

\vspace{1mm}
\noindent \emph{Would it be possible to construct an efficient PSMM/FPMM strategy, simply via designing private queries on top of an SMM strategy?}
\vspace{1mm}

\begin{figure}[htbp]
    \centering
    \includegraphics[scale=0.46]{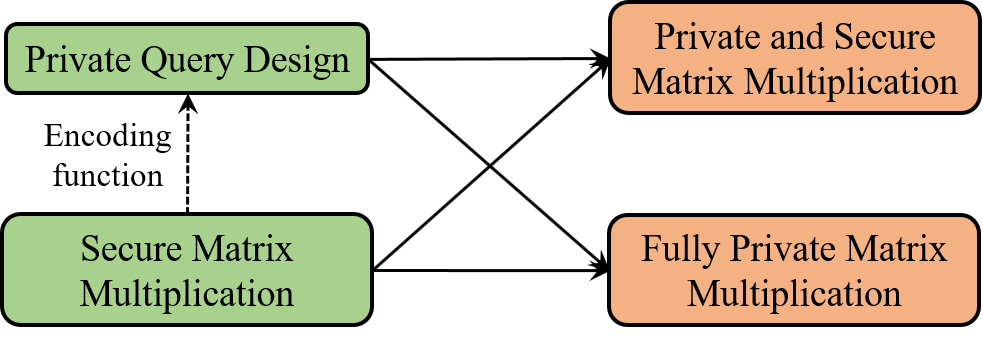}
    \vspace{-2mm}
    \caption{Illustration of the proposed approach to design PSMM and FPMM solutions from a SMM solution and a compatible private query design.}
    \label{SMM_PMM}
\end{figure}

We answer the above question in the affirmative, and propose a novel systematic approach to construct efficient computation strategies for both PSMM and FPMM problems. Specifically, as illustrated in Fig. \ref{SMM_PMM}, we start with an SMM strategy (based on polynomial codes or Lagrange codes), and make use of the special structure inspired by the matrix encoding functions in the SMM strategy to create private queries that facilitate a form of interference alignment, separating the desired and interfering partitions of the matrices in the libraries, such that the response computed at each worker has identical structure as that of the SMM strategy. Consequently, as in the SMM strategy, the desired product can be recovered via polynomial interpolation from the responses, with the same recovery threshold. Our major contributions in this paper are
\begin{itemize}
    \item Establishes a generic connection between the PSMM/FPMM and the SMM problem, which helps to significantly simplify the design process of a PSMM/FPMM strategy;
    \item Compared with state-of-the-art PSMM and FPMM strategies, the strategies constructed from the proposed approach introduce new private query designs, and achieve more flexible tradeoffs between recovery threshold, communication cost and computation complexity, further improving the system efficiency. See Section~\ref{Comparison} for detailed comparisons.
\end{itemize}

\subsection{Related Work}
Coded computing has recently emerged as a technique of utilizing information/coding theoretical tools to inject redundant data and computations into distributed computing systems, to mitigate communication and straggler bottlenecks, and provide security and privacy for various computation tasks (see, e.g.,~\cite{CIT-103,li2015coded,li2017fundamental,li2017scalable,lee2017speeding,li2016unified,dutta2016short,tandon2017gradient,Qian_Yu,MatDot_code,EP_code,LCC,PSMM:Tandon,Tandon_secure_code,Zhu_SDMM,EP_SMC,sun2018capacity,li2020polyshard,zhu2020symmetric}). Privately retrieving a message from a distributed storage system without revealing the index of the message 
has been studied extensively in the problem of Private Information Retrieval (PIR) \cite{chor1995private,shah2014one,Sun_replicated,Sun_TPIR,Ulukus_MDS,Zhu,Zhu_SCPIR} in recent years. 
With the focus on index privacy, the PSMM and FPMM problems
can be viewed as 
secure matrix multiplication problems with additional PIR requirement on the indices of interested matrices within  public libraries.
\subsubsection*{Private and Secure Matrix Multiplication} The problem of private matrix multiplication was first introduced in \cite{PSMM:Kim} without colluding constraint (i.e., $T=1$) and security guarantee on $\mathbf{A}$. The work \cite{PSMM:1} imposed the security constraint on $\mathbf{A}$ to consider a non-colluding PSMM problem, and improved the recovery threshold of the strategy proposed in \cite{PSMM:Kim}, through adopting the random query design in~\cite{shah2014one} to ensure privacy, and employing polynomial codes \cite{Polynomial_code} to complete desired computation.
Subsequently in \cite{PSMM:Tandon}, the authors presented another non-colluding PSMM strategy that combines MDS-coded PIR scheme \cite{Ulukus_MDS} with polynomial codes, and show that the strategy outperforms \cite{PSMM:1} in terms of upload and download communication costs. However, the strategy in~\cite{PSMM:Tandon} provides no resistance to stragglers, and has a high computation complexity.
Further in \cite{MDS-X-security}, for the asymptotic setting (i.e., the number of matrices $V\rightarrow\infty$), a better tradeoff between upload and download cost was achieved by exploiting the idea of PIR based on Cross Subspace Alignment (CSA) \cite{X-security}, at the expense of a higher computation complexity. The authors proposed in \cite{PSMM:2,PSMM:3} novel strategies for the non-colluding PSMM problem using the query design in~\cite{PSMM:1}, yielding a more flexible tradeoff between recovery threshold, communication cost and computation complexity.
Lagrange codes \cite{LCC} were also employed to create non-colluding PSMM strategy \cite{Qian_Yu}, with the help of the query design in \cite{PSMM:1} and bilinear complexity \cite{Smirnov,Strassen}.
A very recent work \cite{PSMM:4} presented a computation strategy based on polynomial codes for the $T$-colluding PSMM problem considered in this paper, but it requires excessive communication cost and computation complexity.

\subsubsection*{Fully Private Matrix Multiplication}
Much less work has been done in the literature for FPMM problem.
In \cite{Qian_Yu}, a non-colluding FPMM strategy was proposed based on Lagrange codes and bilinear complexity, by  resorting to the query design in \cite{PSMM:1}. Later in \cite{FPMM:1}, a $T$-colluding FPMM strategy  was introduced by using the idea of CSA.

In general, the current works \cite{PSMM:2,PSMM:3,Qian_Yu} have well addressed the problems of PSMM and FPMM without colluding constraint. It is valuable to note that, all these works construct their strategies using the query design in \cite{PSMM:1} to ensure privacy, i.e., there is no difference in the private queries sent to workers.
While the  PSMM and FPMM problems with colluding constraint have been studied in \cite{FPMM:1,PSMM:4}, these strategies require either high recovery threshold or huge communication cost and computation complexity. 

\subsection{Organization}
The rest of this paper is organized as follows. In Section \ref{problem statement}, we 
formally formulate the problems of PSMM and FPMM.
In Section \ref{Prrlimi:SMM}, we review the problem of SMM and its strategies based on polynomial codes and Lagrange codes.
In Section \ref{Main result}, we summarize the main results of the paper. Sections \ref{strategy:PSMM} and \ref{section:FPMM} present the proposed computation strategies for PSMM and FPMM, respectively, by exploiting the encoding structure of SMM strategies.
Section \ref{Comparison} gives comparison with other related work.
Finally, the paper is concluded in Section \ref{conclusion}.

\subsubsection*{Notation} Let boldface and cursive capital letters represent matrices and sets, respectively, e.g., $\mathbf{A}$ and $\mathcal{K}$. For a finite set $\mathcal{K}$, $|\mathcal{K}|$ denotes its cardinality.
Denote $\mathbb{Z}^{+}$ the set of positive integers. For any $m,n\in\mathbb{Z}^{+}$ such that $m<n$, $[n]$ and $[m:n]$ denote the sets $\{1,2,\ldots,n\}$ and $\{m,m+1,\ldots,n\}$, respectively.
Define $A_{\mathcal{K}}$ as $\{A_{k_1},\ldots,A_{k_{m}}\}$ for any index set $\mathcal{K}=\{k_1,\ldots,k_{m}\}\subseteq[n]$.

\section{Problem Formulations}\label{problem statement}

Consider a distributed computing system consisting of one master node and $N$ worker nodes, where each worker is connected to the master through an orthogonal communication link. The workers are honest-but-curious, which means that they will follow the prescribed protocol faithfully, yet may potentially collude to infer information about additional data inputs. We consider two private distributed matrix computation problems of \emph{private and secure matrix multiplication} and \emph{fully private matrix multiplication}. In the rest of this section, we describe the formulations of these two problems respectively.

\subsection{Private and Secure Matrix Multiplication}

For the problem of Private and Secure Matrix Multiplication (PSMM) depicted in Fig. \ref{PSMM}, the master owns a \emph{confidential} matrix $\mathbf{A}$ of dimension $\lambda\!\times\!\omega$, and all workers have access to a library $\mathcal{L}^{\mathbf{B}}$ of $V$ \emph{public} matrices $\mathbf{B}^{(1)},\mathbf{B}^{(2)},\ldots,\mathbf{B}^{(V)}$ with dimension $\omega\!\times\!\gamma$, for some $\lambda,\omega,\gamma\in\mathbb{Z}^{+}$.
All the entries of the matrices are over a finite field $\mathbb{F}_q$ for some prime power $q$.

The master privately selects an index $\theta\in[V]$ and wishes to compute the product $\mathbf{A}\mathbf{B}^{(\theta)}$ from the distributed system, while keeping its interested index $\theta$ private and its own matrix $\mathbf{A}$ secure from any colluding subset of up to $T$ out of the $N$ workers. To this end, the master employs a computation strategy of PSMM consisting of the following three phases:
\begin{itemize}
  \item \textbf{Sharing:} To ensure security, the master locally generates a private randomness, denoted by $\mathcal{Z}^{\mathbf{A}}$, which is used to encode the matrix $\mathbf{A}$ according to encoding functions $\bm{f}=(f_1,\ldots,f_{N})$, where $f_i$ is the encoding function for worker $i$. Denote the encoded version of matrix $\mathbf{A}$ for worker $i$ by $\widetilde{\mathbf{A}}_i$, i.e.,
      \begin{IEEEeqnarray}{c}
       \widetilde{\mathbf{A}}_i=f_{i}(\mathbf{A},\mathcal{Z}^{\mathbf{A}}),\quad\forall\,i\in[N].\notag
       \end{IEEEeqnarray}
       To privately complete computation, the master also generates $N$ queries $\mathcal{Q}_{[N]}^{(\theta)}$ based on the index $\theta$ and another locally generated private randomness $\mathcal{Z}^{\theta}$.
       Then the encoded matrix $\widetilde{\mathbf{A}}_i$ and the query $\mathcal{Q}_{i}^{(\theta)}$ are shared with worker $i\in[N]$.
  \item \textbf{Computation:} Upon receiving $\mathcal{Q}_{i}^{(\theta)}$, worker $i$ first uses encoding function $h_i$ to encode the library $\mathcal{L}^{\mathbf{B}}$ into $\widetilde{\mathbf{B}}_i$, i.e.,
    \begin{IEEEeqnarray}{c}\notag
       \widetilde{\mathbf{B}}_i=h_{i}(\mathbf{B}^{([V])},\mathcal{Q}_{i}^{(\theta)}),\quad\forall\,i\in[N],
       \end{IEEEeqnarray}
   and then computes the response $\mathbf{Y}_i^{(\theta)}$ and sends it back to the master, which is a deterministic function of the received $\widetilde{\mathbf{A}}_i$ and the encoded matrix $\widetilde{\mathbf{B}}_i$.
 \item \textbf{Reconstruction:} For some design parameter $K \leq N$, the master only waits for the responses from the fastest $K$ workers, and recovers the desired product $\mathbf{A}\mathbf{B}^{(\theta)}$ from their responses. This allows the computation strategy to tolerate any subset of up to $N-K$ stragglers.
\end{itemize}

A valid PSMM strategy must satisfy the following three requirements.
\begin{itemize}
\item\textbf{Privacy Constraint:} The strategies for computing any two distinct products $\mathbf{A}\mathbf{B}^{(\theta)}$ and $\mathbf{A}\mathbf{B}^{(\theta')}$ must be indistinguishable with respect to any $T$ colluding workers, i.e., for all $\theta\neq\theta'\in[V]$ and $\mathcal{T}\subseteq[N],|\mathcal{T}|=T$,
\begin{IEEEeqnarray*}{c}
(\mathcal{Q}_{\mathcal{T}}^{(\theta)},\widetilde{\mathbf{A}}_{\mathcal{T}},\mathbf{B}^{([V])},\mathbf{Y}_{\mathcal{T}}^{(\theta)})\!\sim\! (\mathcal{Q}_{\mathcal{T}}^{(\theta')},\widetilde{\mathbf{A}}_{\mathcal{T}},\mathbf{B}^{([V])},\mathbf{Y}_{\mathcal{T}}^{(\theta')}),
\end{IEEEeqnarray*}
where $X\sim Y$ means that the random variables $X$ and $Y$ are identically distributed. Equivalently, the index $\theta$ of the desired product is hidden from all the information available to any $T$ colluding workers, i.e.,
\begin{IEEEeqnarray}{c}\label{PSMM:privacy1}
I(\theta;\mathcal{Q}_{\mathcal{T}}^{(\theta)},\widetilde{\mathbf{A}}_{\mathcal{T}},\mathbf{B}^{([V])},\mathbf{Y}_{\mathcal{T}}^{(\theta)})=0.
\end{IEEEeqnarray}
\item\textbf{Security Constraint:} Any $T$ colluding workers must not learn any information about the  confidential matrix $\mathbf{A}$, i.e., for all $\mathcal{T}\subseteq[N],|\mathcal{T}|=T$,
\begin{IEEEeqnarray}{c}
I(\mathbf{A};\mathcal{Q}_{\mathcal{T}}^{(\theta)},\widetilde{\mathbf{A}}_{\mathcal{T}},\mathbf{B}^{([V])},\mathbf{Y}_{\mathcal{T}}^{(\theta)})=0. \label{psmm:privacy}
\end{IEEEeqnarray}
\item\textbf{Correctness Constraint:} The desired product should be correctly reconstructed from the collection of responses of any fastest $K$ workers, i.e.,
\begin{IEEEeqnarray}{c}\notag
       H(\mathbf{A}\mathbf{B}^{(\theta)}|\mathbf{Y}_{\mathcal{K}}^{(\theta)})=0,\quad\forall\,\mathcal{K}\subseteq[N],|\mathcal{K}|=K. 
       \end{IEEEeqnarray}
\end{itemize}

The performance of a PSMM strategy is evaluated by the following key metrics:
\begin{enumerate}
  \item[1.] The recovery threshold $K$, which is the minimum number of workers  that the master needs to wait for in order to recover the desired product $\mathbf{A}\mathbf{B}^{(\theta)}$.
  \item[2.] The communication cost, which is comprised of the upload cost for matrix $\mathbf{A}$ and download cost from workers,\footnote{As in the information-theoretic PIR problem \cite{Sun_replicated,Sun_TPIR,Ulukus_MDS,Zhu,Zhu_SCPIR}, the upload cost for queries can be neglected compared to the upload cost for matrix $\mathbf{A}$ and the download cost, as it does not scale with matrix dimensions. Similarly, in the following, the computation complexity for queries are also neglected.} defined as
    \begin{IEEEeqnarray}{c}
    P_u\!\triangleq\!\frac{\sum_{i=1}^{N}H(\widetilde{\mathbf{A}}_i)}{\lambda\omega}, \; P_d\!\triangleq\!
    \max\limits_{\mathcal{K}:\mathcal{K}\subseteq[N],|\mathcal{K}|=K}\frac{H(\mathbf{Y}_{\mathcal{K}}^{(\theta)})}{\lambda\gamma}, \label{upload and download} \IEEEeqnarraynumspace
    \end{IEEEeqnarray}
    which are normalized with the number of symbols contained in the matrix $\mathbf{A}$ and the desired product $\mathbf{A}\mathbf{B}^{(\theta)}$, respectively.
  \item[3.] The computation complexity, which includes the complexities of encoding, worker computation and decoding. The encoding complexity ${C}_{\mathbf{A}}$ at the master is defined as the number of arithmetic operations required to compute the encoding functions $\bm{f}$. The complexity of worker computation ${C}_{w}$ 
  is defined as the maximal number of arithmetic operations required to compute the response $\mathbf{Y}_i^{(\theta)}$, over all worker $i\in[N]$. Finally, the decoding complexity ${C}_d$ at the master is defined as the 
  maximal number of arithmetic operations required to decode the desired product $\mathbf{A}\mathbf{B}^{(\theta)}$ from the responses of fastest workers in $\mathcal{K}$, over all $\mathcal{K}\subseteq[N]$ with $|\mathcal{K}|=K$.
\end{enumerate}

\subsection{Fully Private Matrix Multiplication}
We describe the problem of Fully Private Matrix Multiplication (FPMM) illustrated in Fig.~\ref{FPMM}. In contrast to the above PSMM problem where the master has a confidential matrix $\mathbf{A}$, in the FPMM problem, there is a library $\mathcal{L}^{\mathbf{A}}$ of $U$ \emph{public} matrices $\mathbf{A}^{(1)},\mathbf{A}^{(2)},\ldots,\mathbf{A}^{(U)}\in\mathbb{F}_q^{\lambda\times\omega}$ that are accessible to the $N$ workers, i.e., each worker has access to the two libraries $\mathcal{L}^{\mathbf{A}}$ and $\mathcal{L}^{\mathbf{B}}$. The master is interested in computing the product $\mathbf{A}^{(\theta_1)}\mathbf{B}^{(\theta_2)}$ utilizing the distributed computing system, while keeping the indices of the desired product $\theta_1$ and $\theta_2$ private from any $T$ colluding workers, for any $\theta_1\in[U]$ and $\theta_2\in[V]$.

To do so, similar to PSMM, a computation strategy for FPMM operates in the following three phases:
\begin{itemize}
  \item \textbf{Sharing:} The master generates the queries $\mathcal{Q}_{[N]}^{(\theta_1)}$ and $\mathcal{Q}_{[N]}^{(\theta_2)}$ for the two libraries $\mathcal{L}^{\mathbf{A}}$ and $\mathcal{L}^{\mathbf{B}}$ according to the interested indices $\theta_1$ and $\theta_2$ and locally generated private randomness $\mathcal{Z}^{\theta_1}$ and $\mathcal{Z}^{\theta_2}$ respectively,
  and then shares $\mathcal{Q}_{i}^{(\theta_1)}$ and $\mathcal{Q}_{i}^{(\theta_2)}$ with worker $i\in[N]$.
  \item \textbf{Computation:} Upon receiving $\mathcal{Q}_{i}^{(\theta_1)}$ and $\mathcal{Q}_{i}^{(\theta_2)}$, worker $i$ first encodes the two libraries $\mathcal{L}^{\mathbf{A}}$ and $\mathcal{L}^{\mathbf{B}}$ using  encoding functions $f_i$ and $h_{i}$, respectively. The encoded versions of $\mathcal{L}^{\mathbf{A}}$ and $\mathcal{L}^{\mathbf{B}}$ for worker $i\in[N]$ are given by
      \begin{IEEEeqnarray}{c}\notag
       \widetilde{\mathbf{A}}_i=f_{i}(\mathbf{A}^{([U])},\mathcal{Q}_{i}^{(\theta_1)}),\quad\widetilde{\mathbf{B}}_i=h_{i}(\mathbf{B}^{([V])},\mathcal{Q}_{i}^{(\theta_2)}).
       \end{IEEEeqnarray}
  Then the worker $i$ computes the response $\mathbf{Y}_i^{(\theta_1,\theta_2)}$ and sends it back to the master, which is a function of the encoded matrices $\widetilde{\mathbf{A}}_i$ and $\widetilde{\mathbf{B}}_i$.

 \item \textbf{Reconstruction:} The master recovers the desired product $\mathbf{A}^{(\theta_1)}\mathbf{B}^{(\theta_2)}$ from the responses of any fastest $K$ workers.
\end{itemize}

A valid computation strategy for FPMM must satisfy the following constraints.
\begin{itemize}
\item\textbf{Fully Privacy Constraint:} The desired indices $\theta_1$ and $\theta_2$ must be hidden from all the information available to any $T$ colluding workers, i.e., for all $\mathcal{T}\subseteq[N],|\mathcal{T}|=T$,
\begin{IEEEeqnarray}{c}\label{FPMM:privacy:235}
I(\theta_1,\theta_2;\mathcal{Q}_{\mathcal{T}}^{(\theta_1)},\mathcal{Q}_{\mathcal{T}}^{(\theta_2)},\mathbf{A}^{([U])},\mathbf{B}^{([V])},\mathbf{Y}_{\mathcal{T}}^{(\theta_1,\theta_2)})=0. \IEEEeqnarraynumspace
\end{IEEEeqnarray}
\item\textbf{Correctness Constraint:} With the responses of any fastest $K$ workers, the desired product must be recovered, i.e.,
\begin{IEEEeqnarray}{c}\notag
       H(\mathbf{A}^{(\theta_1)}\mathbf{B}^{(\theta_2)}|\mathbf{Y}_{\mathcal{K}}^{(\theta_1,\theta_2)})=0,\quad\forall\,\mathcal{K}\subseteq[N],|\mathcal{K}|=K.
       \end{IEEEeqnarray}
\end{itemize}

Similar to the PSMM problem, the performance of an FPMM strategy is evaluated by the following key quantities: 1) the recovery threshold $K$; 2) the normalized download cost $P_d$; and 3) the computation complexities consisting of generating response at each worker ${C}_w$ and decoding desired product at the master ${C}_d$.

For the above formulated PSMM and FPMM problems, our goal in this paper is to design efficient computation strategies that minimize the recovery threshold, the communication cost and the computation complexity. As the first step, we review the strategies proposed to solve a related distributed matrix multiplication problem, which serves as a building block of our approach to solve the private matrix multiplication problems.

\section{Preliminaries: Secure Matrix Multiplication}\label{Prrlimi:SMM}
In this section, we summarize the state-of-the-art computation strategies for the Secure Matrix Multiplication (SMM) problem, which will be exploited to construct the strategies for PSMM and FPMM in the following sections. In the SMM problem, the master owns two confidential matrices $\mathbf{A}\in\mathbb{F}_q^{\lambda\times\omega}$ and $\mathbf{B}\in\mathbb{F}_q^{\omega\times\gamma}$, and is interested in computing the product $\mathbf{C}=\mathbf{AB}$ in the distributed computing system, without revealing anything about $\mathbf{A}$ and $\mathbf{B}$ to any $T$ colluding workers.

We first introduce a lemma that will be used in security and privacy proofs.
\begin{Lemma}[Generalized Secret Sharing \cite{Shamir,Zhu_SDMM}]\label{security proof}
For any parameters $L,T,\kappa,\tau\in\mathbb{Z}^{+}$, let $\mathbf{W}_{1},\ldots,\mathbf{W}_{L}\in\mathbb{F}_q^{\kappa\times\tau}$ be $L$ secrets, and $\mathbf{Z}_{1},\ldots,\mathbf{Z}_{T}$ be $T$ random matrices with the same dimensions as secrets whose entries are chosen independently and uniformly from $\mathbb{F}_q$. Let $\alpha_1,\ldots,\alpha_N$ be $N$ pairwise distinct elements from $\mathbb{F}_q$.
Define a function of $x$ as
\begin{IEEEeqnarray}{c}\notag
y(x)\!=\!\mathbf{W}_{\!1}u_1(x)\!+\!\ldots\!+\!\mathbf{W}_{\!L}u_L(x)\!+\!\mathbf{Z}_{1}v_1(x)\!+\!\ldots\!+\!\mathbf{Z}_{T}v_T(x),
\end{IEEEeqnarray}
where $u_1(x),\ldots,u_L(x),v_1(x),\ldots,v_T(x)\in\mathbb{F}_q[x]$ are the deterministic functions of $x$.
If the matrix
\begin{IEEEeqnarray}{c}\notag
\mathbf{V}=
\left[
  \begin{array}{@{}ccc@{}}
    v_1(\alpha_{i_1})  & \ldots & v_T(\alpha_{i_1}) \\
    \vdots & \ddots & \vdots \\
    v_1(\alpha_{i_T})  & \ldots & v_T(\alpha_{i_T}) \\
  \end{array}
\right]_{T\times T}
\end{IEEEeqnarray}
is non-singular over $\mathbb{F}_q$ for any $\mathcal{T}=\{i_1,\ldots,i_T\}\subseteq[N]$ with $|\mathcal{T}|=T$, then the $T$ values $y(\alpha_{i_1}),\ldots,y(\alpha_{i_T})$  can not learn any information about the secrets $\mathbf{W}_{1},\ldots,\mathbf{W}_{L}$, i.e.,
\begin{IEEEeqnarray}{c}\notag
I(y(\alpha_{i_1}),\ldots,y(\alpha_{i_T});\mathbf{W}_{1},\ldots,\mathbf{W}_{L})=0.
\end{IEEEeqnarray}
\end{Lemma}

Let $m,p,n\in\mathbb{Z}^{+}$ be any partitioning parameters of data matrices such that $m|\lambda,p|\omega$ and $n|\gamma$.  To efficiently exploit the computation power of distributed workers and establish the feasible tradeoff between system performance, the matrices $\mathbf{A}$ and $\mathbf{B}$ are partitioned into $m\times p$ and $p\times n$ equal-size sub-matrices, respectively, as shown below.
\begin{IEEEeqnarray}{c}\label{partition3214}
\mathbf{A}\!=\!
\left[
  \begin{array}{@{}ccc@{}}
    \mathbf{A}_{1,1}  & \ldots & \mathbf{A}_{1,p} \\
    \vdots & \ddots & \vdots \\
    \mathbf{A}_{m,1}  & \ldots & \mathbf{A}_{m,p} \\
  \end{array}
\right], \quad
\mathbf{B}\!=\!
\left[
  \begin{array}{@{}ccc@{}}
    \mathbf{B}_{1,1} & \ldots & \mathbf{B}_{1,n} \\
    \vdots  & \ddots & \vdots \\
    \mathbf{B}_{p,1}  & \ldots & \mathbf{B}_{p,n} \\
  \end{array}
\right], \IEEEeqnarraynumspace
\end{IEEEeqnarray}
where $\mathbf{A}_{k,\ell}\in\mathbb{F}_q^{\frac{\lambda}{m}\times\frac{\omega}{p}}$ for any $k\in[m],\ell\in[p]$ and $\mathbf{B}_{\ell,j}\in\mathbb{F}_q^{\frac{\omega}{p}\times\frac{\gamma}{n}}$ for any $\ell\in[p],j\in[n]$.
Accordingly, the desired product $\mathbf{C}=\mathbf{AB}$ involves a total of $mn$ linear combinations of products of sub-matrices, i.e.,
\begin{IEEEeqnarray}{c}\label{EP code Result}
\mathbf{C}=\mathbf{AB}=
\left[
  \begin{array}{@{}ccc@{}}
    \mathbf{C}_{1,1} & \ldots & \mathbf{C}_{1,n} \\
    \vdots & \ddots & \vdots \\
    \mathbf{C}_{m,1}  & \ldots & \mathbf{C}_{m,n} \\
  \end{array}
\right],
\end{IEEEeqnarray}
where $\mathbf{C}_{k,j}=\sum_{\ell=1}^{p}\mathbf{A}_{k,\ell}\mathbf{B}_{\ell,j}$ for all $k\in[m],j\in[n]$.

To the best of our knowledge, the state-of-the-art strategies for SMM with arbitrary partitioning of matrices above can be divided into two categories in terms of coding techniques, i.e., SMM based on polynomial codes \cite{Zhu_SDMM,EP_SMC} and SMM based on Lagrange codes \cite{Qian_Yu}. The essential components behind these coded strategies lie in constructing the encoding functions of matrices $\mathbf{A}$ and $\mathbf{B}$, denoted by $f(x)$ and $h(x)$ respectively, such that the desired products of sub-matrices $\mathbf{C}_{k,j},k\in[m],j\in[n]$ can be recovered by interpolating the product polynomial $g(x)=f(x)\cdot h(x)$. 
Next, we present these two approaches and their performance.

\subsection{Secure Matrix Multiplication Based on Polynomial Codes}\label{SMM:PC}
Let $\mathbf{Z}_{1}^{\mathbf{A}},\ldots,\mathbf{Z}_{T}^{\mathbf{A}}$ and $\mathbf{Z}_{1}^{\mathbf{B}},\ldots,\mathbf{Z}_{T}^{\mathbf{B}}$ be $T$ random matrices distributed independently and uniformly on $\mathbb{F}_q^{\frac{\lambda}{m}\times\frac{\omega}{p}}$ and $\mathbb{F}_q^{\frac{\omega}{p}\times\frac{\gamma}{n}}$, respectively.
In general, the goal of SMM strategies based on polynomial codes is to design a group of appropriate positive integers $\{a_{k,\ell},b_{\ell,j},c_t,d_t:k\in[m],\ell\in[p],j\in[n],t\in[T]\}$ to construct the encoding functions of matrices $\mathbf{A}$ and $\mathbf{B}$ \eqref{partition3214} as
\begin{IEEEeqnarray}{rCl}
f_P(x)&=&\sum\limits_{k=1}^{m}\sum\limits_{\ell=1}^{p}\mathbf{A}_{k,\ell}x^{a_{k,\ell}}+\sum\limits_{t=1}^{T}\mathbf{Z}_{t}^{\mathbf{A}}x^{c_t}, \label{encoded marix:C1}\\
h_P(x)&=&\sum\limits_{\ell=1}^{p}\sum\limits_{j=1}^{n}\mathbf{B}_{\ell,j}x^{b_{\ell,j}}+\sum\limits_{t=1}^{T}\mathbf{Z}_{t}^{\mathbf{B}}x^{d_t}, \label{encoded marix:C2}
\end{IEEEeqnarray}
such that the following criteria are satisfied:
\begin{itemize}
\item[C1] The product polynomial
\begin{IEEEeqnarray}{c}\notag
g(x)=\sum\limits_{r=0}^{\delta}g_r x^r=f_P(x)\cdot h_P(x)
\end{IEEEeqnarray}
contains all the desired sub-products $\mathbf{C}_{k,j},k\in[m],j\in[n]$ as coefficients, i.e., $\mathbf{C}_{k,j}\in\{g_r:r\in[0:\delta]\}$ for all $k\in[m],j\in[n]$, where $g_r$ is the coefficient of $x^r$ in $f_P(x)\cdot h_P(x)$, and $\delta$ is the degree of the polynomial $f_P(x)\cdot h_P(x)$, given by
\begin{IEEEeqnarray}{l}
\delta\triangleq\max\{a_{k,\ell},c_t:k\in[m],\ell\in[p],t\in[T]\}\notag\\
\quad\quad\quad\quad\quad+\max\{b_{\ell,j},d_t:\ell\in[p],j\in[n],t\in[T]\}. \notag
\end{IEEEeqnarray}
\item[C2] The following matrices
\begin{IEEEeqnarray}{c}\label{SMM:C2}
\left[
  \begin{array}{@{}c@{\;}c@{\;}c@{\;}c@{}}
    \alpha_{i_1}^{c_1} & \alpha_{i_1}^{c_2} & \ldots & \alpha_{i_1}^{c_T} \\
    \alpha_{i_2}^{c_1} & \alpha_{i_2}^{c_2} & \ldots & \alpha_{i_2}^{c_T} \\
    \vdots & \vdots & \ddots & \vdots \\
    \alpha_{i_T}^{c_1} & \alpha_{i_T}^{c_2} & \ldots & \alpha_{i_T}^{c_T} \\
  \end{array}
\right]_{T\times T}, \;
\left[
  \begin{array}{@{}c@{\;}c@{\;}c@{\;}c@{}}
    \alpha_{i_1}^{d_1} & \alpha_{i_1}^{d_2} & \ldots & \alpha_{i_1}^{d_T} \\
    \alpha_{i_2}^{d_1} & \alpha_{i_2}^{d_2} & \ldots & \alpha_{i_2}^{d_T} \\
    \vdots & \vdots & \ddots & \vdots \\
    \alpha_{i_T}^{d_1} & \alpha_{i_T}^{d_2} & \ldots & \alpha_{i_T}^{d_T} \\
  \end{array}
\right]_{T\times T} \IEEEeqnarraynumspace
\end{IEEEeqnarray}
are non-singular over $\mathbb{F}_q$ for any $\mathcal{T}=\{i_1,\ldots,i_T\}\subseteq[N]$ with $|\mathcal{T}|=T$, where $\alpha_1,\alpha_2,\ldots,\alpha_N$ are $N$ pairwise distinct non-zero elements from $\mathbb{F}_q$.
\end{itemize}

To complete the computation $\mathbf{C}=\mathbf{AB}$ \eqref{EP code Result}, the master first shares the evaluations of $f_P(x)$ and $h_P(x)$ at $x=\alpha_i$ with worker $i\in[N]$. Then the worker $i$ computes the product $f_P(\alpha_i)\cdot h_P(\alpha_i)$ and sends it back to the master on successful completion, which is equivalent to evaluating of the polynomial $g(x)=f_P(x)\cdot h_P(x)$ at $x=\alpha_i$. Thus, the master can interpolate $g(x)$ from any $K=\delta+1$ responses by using Lagrange interpolation rule, and then recovers all the desired products of sub-matrices $\mathbf{C}_{k,j},k\in[m],j\in[n]$ from the coefficients of $g(x)$ by C1. The matrices $\mathbf{A}$ and $\mathbf{B}$ are secure against any $T$ colluding workers by C2 and Lemma \ref{security proof}.
Therefore, the SMM strategy based on polynomial codes achieves the recovery threshold $K=\delta+1$.

As far as we know, the state-of-the-art strategies for SMM based on polynomial codes are reflected in \cite{Zhu_SDMM,EP_SMC}, which are summarized in the following lemma.
\begin{Lemma}\label{current scheme:polynomial codes}
For any arbitrary partitioning of matrices $\mathbf{A}$ and $\mathbf{B}$ with parameters $m,p,n$, the state-of-the-art strategies for SMM based on polynomial codes achieve
\begin{itemize}
  \item the recovery threshold $K=(m+1)(np+T)-1$ \cite{Zhu_SDMM} by setting $a_{k,\ell}=(k-1)(np+T)+\ell-1,b_{\ell,j}=jp-\ell,c_t=(m-1)(np+T)+np+t-1,d_t=np+t-1$ for all $k\in[m],\ell\in[p],j\in[n],t\in[T]$,
  \item the recovery threshold $K=(n+1)(mp+T)-1$ \cite{Zhu_SDMM} by setting $a_{k,\ell}=(k-1)p+\ell-1,b_{\ell,j}=(j-1)(mp+T)+p-\ell,c_t=mp+t-1,d_t=(n-1)(mp+T)+mp+t-1$ for all $k\in[m],\ell\in[p],j\in[n],t\in[T]$,
  \item and the recovery threshold $K=2mpn+2T-1$ \cite{EP_SMC} by setting $a_{k,\ell}=(k-1)np+\ell-1,b_{\ell,j}=jp-\ell,c_t=mpn+t-1,d_t=mpn+t-1$ for all $k\in[m],\ell\in[p],j\in[n],t\in[T]$.
\end{itemize}
That is, the current best SMM strategies based on polynomial codes achieve the recovery threshold $K=\min\{(m+1)(np+T)-1,(n+1)(mp+T)-1,2mpn+2T-1\}$.
\end{Lemma}

\subsection{Secure Matrix Multiplication Based on Lagrange Codes}\label{SMM:LC}
Lagrange codes were originally introduced in \cite{LCC} and is widely applied to solve batch processing problems for coded distributed computing \cite{MatDot_code,batch_matrix}. To solve SMM using Lagrange codes \cite{EP_code,Qian_Yu}, matrix multiplication is first converted into the problem of computing the element-wise product of two batches of sub-matrices, by employing the concept of bilinear complexity, and then Lagrange codes can directly operate on the problem of batch sub-matrix multiplication.

\begin{Definition}[Bilinear Complexity \cite{Strassen,Smirnov}]\label{Def Bilinear Complexity}
Let $\mathbf{C}=[C_{k,j}]_{k\in[m],j\in[n]}$ be the product of any matrices $\mathbf{A}=[A_{k,\ell}]_{k\in[m],\ell\in[p]}$ and $\mathbf{B}=[B_{\ell,j}]_{\ell\in[p],j\in[n]}$, where $C_{k,j}=\sum_{\ell=1}^{p}A_{k,\ell}B_{\ell,j}$.
The bilinear complexity, denoted by $R(m,p,n)$, is defined as the minimum number of active multiplications for the problem of multiplying the two matrices $\mathbf{A}$ and $\mathbf{B}$. Moreover, an upper bound construction with rank $R$ for bilinear complexity\footnote{The upper bound construction of bilinear complexity is known for many cases of parameters $m,p,n$, based on the recursive method in \cite{Smirnov}.} means that, there exists tensors $a\in\mathbb{F}_q^{R\times m \times p},b\in\mathbb{F}_q^{R\times p \times n},c\in\mathbb{F}_q^{R\times m \times n}$ satisfying
\begin{IEEEeqnarray*}{l}
\sum\limits_{r=1}^{R}c_{r,k,j}\Bigg(\underbrace{\sum\limits_{k'=1}^{m}\sum\limits_{\ell'=1}^{p}a_{r,k',\ell'}A_{k',\ell'}}_{= A_r} \Bigg) \Bigg(\underbrace{\sum\limits_{\ell'=1}^{p}\sum\limits_{j'=1}^{n}b_{r,\ell',j'}B_{\ell',j'}}_{= B_r} \Bigg)\notag\\
\quad\quad\quad\quad\quad\quad=\sum\limits_{\ell=1}^{p}A_{k,\ell}B_{\ell,j}\!=\!C_{k,j},\quad\forall\,k\in[m],j\in[n].\label{dec Bilinear Complexity}
\end{IEEEeqnarray*}
\end{Definition}

It is straightforward to observe from Definition \ref{Def Bilinear Complexity} that, bilinear complexity enables converting the matrix multiplication problem $\mathbf{C}=\mathbf{AB}$ in  \eqref{EP code Result} into computing the element-wise products of two batches of sub-matrices of length $R$. Specifically, given any upper bound construction for bilinear complexity with tensors $a=(a_{r,k,\ell}),b=(b_{r,\ell,j}),c=(c_{r,k,j})$ and rank $R$, the desired products of sub-matrices $\mathbf{C}_{k,j}$ can be recovered by
\begin{IEEEeqnarray}{c}
\sum\limits_{r=1}^{R}c_{r,k,j}\mathbf{A}_r\mathbf{B}_r\!=\!\sum\limits_{\ell=1}^{p}\mathbf{A}_{k,\ell}\mathbf{B}_{\ell,j}\!=\!\mathbf{C}_{k,j},\;\forall\,k\!\in\![m],j\!\in\![n],\label{lag:recover} \IEEEeqnarraynumspace
\end{IEEEeqnarray}
if one obtains the element-wise product $\{\mathbf{A}_r\mathbf{B}_r:r\in[R]\}$ of the two batches of sub-matrices $(\mathbf{A}_1,\mathbf{A}_2,\ldots,\mathbf{A}_{R})$ and $(\mathbf{B}_1,\mathbf{B}_2,\ldots,\mathbf{B}_{R})$, where
\begin{IEEEeqnarray}{cl}
\mathbf{A}_{r}=\sum\limits_{k=1}^{m}\sum\limits_{\ell=1}^{p}a_{r,k,\ell}\mathbf{A}_{k,\ell},& \quad\forall\, r\in[R],  \label{bilinear:constru:1}\\
\mathbf{B}_{r}=\sum\limits_{\ell=1}^{p}\sum\limits_{j=1}^{n}b_{r,\ell,j}\mathbf{B}_{\ell,j}, &\quad\forall\, r\in[R]. \label{bilinear:constru:2}
\end{IEEEeqnarray}
That is, bilinear complexity converts the matrix multiplication $\mathbf{C}=\mathbf{AB}$ into the problem of computing the element-wise product of two batches of sub-matrices $(\mathbf{A}_1,\mathbf{A}_2,\ldots,\mathbf{A}_{R})$ and $(\mathbf{B}_1,\mathbf{B}_2,\ldots,\mathbf{B}_{R})$. Then Lagrange codes are operated as follows.

Let $\{\beta_{r},\alpha_i:r\in[R+T],i\in[N]\}$ be $R+T+N$ distinct elements from $\mathbb{F}_q$.
Construct the encoding functions $f_L(x)$ and $h_L(x)$ of the two batches of sub-matrices as Lagrange interpolation polynomials of degree $R+T-1$, such that
\begin{IEEEeqnarray}{rCl}
f_L(\beta_{r})&=&\left\{
\begin{array}{@{}ll}
\mathbf{A}_{r},&\forall\, r\in[R]\\
\mathbf{Z}_{r}^{\mathbf{A}},&\forall\, r\in[R+1:R+T]
\end{array}\right. \label{bilinear:1}\\
h_L(\beta_{r})&=&\left\{
\begin{array}{@{}ll}
\mathbf{B}_{r},&\forall\, r\in[R]\\
\mathbf{Z}_{r}^{\mathbf{B}},&\forall\, r\in[R+1:R+T]
\end{array}\right., \label{bilinear:2}
\end{IEEEeqnarray}
where $\mathbf{Z}_{R+1}^{\mathbf{A}},\ldots,\mathbf{Z}_{R+T}^{\mathbf{A}}$ and $\mathbf{Z}_{R+1}^{\mathbf{B}},\ldots,\mathbf{Z}_{R+T}^{\mathbf{B}}$ are random matrices over $\mathbb{F}_q$ with the same dimensions as $\mathbf{A}_{r}$ and $\mathbf{B}_{r}$, respectively.
By Lagrange interpolation rule, the polynomials $f_L(x)$ and $h_L(x)$ are written as
\begin{IEEEeqnarray}{rCl}
f_L(x)&=&\sum\limits_{r=1}^{R}\mathbf{A}_{r}\cdot\prod_{j\in[R+T]\backslash\{r\}}\frac{x-\beta_{j}}{\beta_{r}-\beta_{j}}\notag\\
&&\quad\quad\quad+\sum\limits_{k=R+1}^{R+T}\mathbf{Z}_{k}^{\mathbf{A}}\cdot\prod_{j\in[R+T]\backslash\{k\}}\frac{x-\beta_{j}}{\beta_{k}-\beta_{j}},\label{encoding funcion:lagrange:A}\IEEEeqnarraynumspace\\
h_L(x)&=&\sum\limits_{r=1}^{R}\mathbf{B}_{r}\cdot\prod_{j\in[R+T]\backslash\{r\}}\frac{x-\beta_{j}}{\beta_{r}-\beta_{j}}\notag\\
&&\quad\quad\quad+\sum\limits_{k=R+1}^{R+T}\mathbf{Z}_{k}^{\mathbf{B}}\cdot\prod_{j\in[R+T]\backslash\{k\}}\frac{x-\beta_{j}}{\beta_{k}-\beta_{j}}. \label{encoding funcion:lagrange:B} \IEEEeqnarraynumspace
\end{IEEEeqnarray}

To complete computation, the master first shares the evaluations of $f_L(x)$ and $h_L(x)$ at point $x=\alpha_i$ with worker $i\in[N]$, who then responds with the product $f_L(\alpha_i)\cdot h_L(\alpha_i)$. Apparently, the master can interpolate the product polynomial $g(x)=f_L(x)\cdot h_L(x)$ from any $K=\deg(g(x))+1=2R+2T-1$ responses, and then evaluates $g(x)$ at $x=\beta_1,\beta_2,\ldots,\beta_R$ to obtain the element-wise product $\{\mathbf{A}_r\mathbf{B}_r:r\in[R]\}$ by \eqref{bilinear:1}-\eqref{bilinear:2}. It is straightforward to prove the security of $\mathbf{A}$ and $\mathbf{B}$ by Lemma \ref{security proof}. Hence, the SMM strategy based on Lagrange codes achieves the recovery threshold $K=2R+2T-1$.

\section{Main Results}\label{Main result}
We strate our main results in this section. For brevity, we focus on the results of recovery threshold, and present the communication costs and computation complexities of proposed strategies for PSMM (resp. FPMM) in Section \ref{proof:strategy} (resp. \ref{proof:strategy:2}).

\begin{Theorem}\label{theorem:PSMM:PC}
For the problem of private and secure matrix multiplication with $T$ colluding workers and the partitioning parameters $m,p,n$, give any positive integers $\{a_{k,\ell},b_{\ell,j},c_t,d_t:k\in[m],\ell\in[p],j\in[n],t\in[T]\}$ satisfying C1-C2, there exits a computation strategy based on polynomial codes that achieves a recovery threshold of $K=\max\{a_{k,\ell},c_t:k\in[m],\ell\in[p],t\in[T]\}+\max\{b_{\ell,j},d_t:\ell\in[p],j\in[n],t\in[T]\}+1$.
\end{Theorem}
Theorem \ref{theorem:PSMM:PC} is proved in Section  \ref{section:PSMM:PC} by presenting our proposed PSMM strategy based on polynomial codes. The confidential matrix ${\bf A}$ and index $\theta$ are secured by polynomial codes such that the response computed at each worker resembles that of the SMM strategy in Section~\ref{SMM:PC}.

From Lemma \ref{current scheme:polynomial codes}, we can directly obtain the following corollary of Theorem~\ref{theorem:PSMM:PC}.

\begin{Corollary}\label{cor:PSMM-poly}
The recovery threshold $K=\min\{(m+1)(np+T)-1,(n+1)(mp+T)-1,2mpn+2T-1\}$ can be achieved by some PSMM strategy based on polynomial codes.
\end{Corollary}

\begin{Remark}
While the recovery threshold presented in Corollary~\ref{cor:PSMM-poly} is obtained by adopting the degree parameters in Lemma~\ref{current scheme:polynomial codes} for the underlying polynomial codes, we can flexibly optimize system performance over degree parameters $\{a_{k,\ell},b_{\ell,j},c_t,d_t:k\in[m],\ell\in[p],j\in[n],t\in[T]\}$ that satisfy conditions C1-C2. This remains to be an interesting future research problem.
\end{Remark}

\begin{Theorem}\label{theorem:PSMM:LC}
For the problem of private and secure matrix multiplication with $T$ colluding workers and the partitioning parameters $m,p,n$, there exits a computation strategy based on Lagrange codes that achieves a recovery threshold of
$K=2R+2T-1$, where $R$ denotes the rank of any construction for bilinear complexity of multiplying an $m$-by-$p$ matrix and a $p$-by-$n$ matrix.
\end{Theorem}

Theorem \ref{theorem:PSMM:LC} is proved in Section \ref{section:PSMM:LC} by constructing a PSMM strategy based on Lagrange codes. The confidential matrix ${\bf A}$ and index $\theta$ are secured by Lagrange codes such that the response computed at each worker resembles that of the SMM strategy in Section~\ref{SMM:LC}.

\begin{Remark}\label{PSMM:remark}
As far as we know, the general upper bound construction $R$ for bilinear complexity remains open.
Reference \cite{Sedoglavic} lists the current best known upper construction of bilinear complexity for almost all possible partitioning parameters $m,p,n$ with $m\in[2:32],m\leq p\leq n\leq 32$.
For some specific combinations of  parameters $(m,p,n)$, we compare the recovery thresholds achieved by our proposed PSMM strategies in Table \ref{tab:comparative22}. For each specific combination of $(m,p,n)$, which strategy achieves a smaller recovery threshold depends on the value of the security parameter $T$. For example, for the parameter case of $(m=3,p=3,n=3)$, the PSMM strategy based on polynomial codes outperforms the one based on Lagrange codes in terms of recovery threshold when $T<5$. A similar discussion holds for the following FPMM strategies based on polynomial codes and Lagrange codes.
The detailed comparisons between the two proposed strategies for PSMM and FPMM problems  are presented in Sections \ref{com:PSMM:2} and \ref{FPMM:COM:2}, respectively.

\begin{table*}[htbp]
\centering
\caption{Recovery thresholds of the proposed PSMM strategies based on polynomial codes and Lagrange codes.}
  \begin{tabular}{|c|c|c|c|}
  \hline
  \multirow{2}*{\minitab[c]{Partitioning \\ Parameters}} & \multirow{2}*{\minitab[c]{Best Known Bilinear \\ Complexity $R$ \cite{Sedoglavic}}}  & \multicolumn{2}{c|}{Recovery Threshold for PSMM}  \\ \cline{3-4}
   &  & Polynomial codes Based & Lagrange codes Based \\ \hline
  $m=2,p=2,n=2$ & $7$  & $\min\{11+3T,15+2T\}$ & $13+2T$  \\ \hline
  $m=3,p=3,n=3$ & $23$  & $\min\{35+4T,53+2T\}$ & $45+2T$   \\ \hline
  $m=5,p=5,n=5$ & $98$  & $\min\{149+6T,249+2T\}$ & $195+2T$  \\ \hline
  \end{tabular}
  \label{tab:comparative22}
\end{table*}
\end{Remark}

We next turn to present the results of recovery thresholds achieved by our proposed strategies, for the fully private matrix multiplication problem.

\begin{Theorem}\label{theorem:FPMM:1}
For the problem of fully private matrix multiplication with $T$ colluding workers and the partitioning parameters $m,p,n$, give any positive integers $\{a_{k,\ell},b_{\ell,j},c_t,d_t:k\in[m],\ell\in[p],j\in[n],t\in[T]\}$ satisfying C1-C2, there exists a computation strategy based on polynomial codes that achieves a recovery threshold of $K=\max\{a_{k,\ell},c_t:k\in[m],\ell\in[p],t\in[T]\}+\max\{b_{\ell,j},d_t:\ell\in[p],j\in[n],t\in[T]\}+1$.
\end{Theorem}
Theorem \ref{theorem:FPMM:1} is proved in Section \ref{proof:theorem:FPMM} by presenting a FPMM strategy based on polynomial codes. The confidential matrix indices $\theta_1$ and $\theta_2$ are secured by polynomial codes such that the response computed at each worker resembles that of the SMM strategy in Section~\ref{SMM:PC}.

The following corollary is immediate from Lemma \ref{current scheme:polynomial codes}.
\begin{Corollary}
The recovery threshold $K=\min\{(m+1)(np+T)-1,(n+1)(mp+T)-1,2mpn+2T-1\}$ can be achieved by some FPMM strategy based on polynomial codes.
\end{Corollary}

\begin{Theorem}\label{theorem:FPMM:2}
For the problem of fully private matrix multiplication with $T$ colluding workers and the partitioning parameters $m,p,n$, there exists a computation strategy based on Lagrange codes that achieves a recovery threshold of
$K=2R+2T-1$, where $R$ denotes the rank of any construction for bilinear complexity of multiplying an $m$-by-$p$ matrix and a $p$-by-$n$ matrix.
\end{Theorem}

Theorem \ref{theorem:FPMM:2} is proved in Section \ref{proof:theorem:FPMM2} by presenting a FPMM strategy based on Lagrange codes. The confidential matrix indices $\theta_1$ and $\theta_2$ are secured by Lagrange codes such that the response computed at each worker resembles that of the SMM strategy in Section~\ref{SMM:LC}.

\begin{Remark}
We may further consider the presence of some adversarial workers of size $E$ who maliciously return arbitrarily erroneous responses to the master. In our proposed strategies for PSMM and FPMM, the responses of all the workers can be viewed as evaluations of a polynomial at distinct points, and accordingly the responses constitute a Reed-Solomon codeword. Thus, our proposed strategies can provide robustness against the $E$ adversarial workers by waiting for responses from $2E$ more workers.
\end{Remark}

\section{Computation Strategies for Private and Secure Matrix Multiplication}\label{strategy:PSMM}
In this section, we first present two PSMM strategies, which are constructed by exploiting the structure of SMM strategies based on polynomial codes and Lagrange codes, respectively. Then their security, privacy, communication cost and computation complexities are analysed.
This provides proofs for Theorems \ref{theorem:PSMM:PC} and \ref{theorem:PSMM:LC}.

To better attain the tradeoff with respect to system performance, similar to \eqref{partition3214}, the matrices $\mathbf{A}$ and $\mathbf{B}^{(v)}$ are divided into $m\times p$ and $p\times n$ equal-size sub-matrices, respectively, for  any partitioning parameters $m,p,n$, i.e., for all $v\in[V]$,
\begin{IEEEeqnarray}{c}\label{GT1:partion1}
\mathbf{A}\!=\!
\left[
  \begin{array}{@{}c@{\;\,}c@{\;\,}c@{}}
    \mathbf{A}_{1,1}  & \ldots & \mathbf{A}_{1,p} \\
    \vdots  & \ddots & \vdots \\
    \mathbf{A}_{m,1}  & \ldots & \mathbf{A}_{m,p} \\
  \end{array}
\right],\;\;
\mathbf{B}^{(v)}\!=\!
\left[
  \begin{array}{@{}c@{\;\,}c@{\;\,}c@{}}
    \mathbf{B}_{1,1}^{(v)} &  \ldots & \mathbf{B}_{1,n}^{(v)} \\
    \vdots & \ddots & \vdots \\
    \mathbf{B}_{p,1}^{(v)}  & \ldots & \mathbf{B}_{p,n}^{(v)} \\
  \end{array}
\right], \IEEEeqnarraynumspace
\end{IEEEeqnarray}
where $\mathbf{A}_{k,\ell}\in\mathbb{F}_q^{\frac{\lambda}{m}\times\frac{\omega}{p}}$ for any $k\in[m],\ell\in[p]$, and $\mathbf{B}_{\ell,j}^{(v)}\in\mathbb{F}_q^{\frac{\omega}{p}\times\frac{\gamma}{n}}$ for any $\ell\in[p],j\in[n]$.
Then the desired product $\mathbf{C}^{(\theta)}=\mathbf{A}\mathbf{B}^{(\theta)}$ is given by
\begin{IEEEeqnarray}{c}\notag
\mathbf{C}^{(\theta)}=\mathbf{A}\mathbf{B}^{(\theta)}=
\left[
  \begin{array}{@{}ccc@{}}
    \mathbf{C}_{1,1}^{(\theta)} & \ldots & \mathbf{C}_{1,n}^{(\theta)} \\
    \vdots & \ddots & \vdots \\
    \mathbf{C}_{m,1}^{(\theta)} & \ldots & \mathbf{C}_{m,n}^{(\theta)} \\
  \end{array}
\right]
\end{IEEEeqnarray}
with $\mathbf{C}_{k,j}^{(\theta)}=\sum_{\ell=1}^{p}\mathbf{A}_{k,\ell}\mathbf{B}_{\ell,j}^{(\theta)}$ for any $k\in[m],j\in[n]$.

\subsection{PSMM Strategy Based on Polynomial Codes}\label{section:PSMM:PC}
We start with proving Theorem \ref{theorem:PSMM:PC}. We show that any SMM strategy based on polynomial codes can be exploited to construct a PSMM strategy with same recovery threshold. First, we illustrate the key idea behind the proposed PSMM strategy through a simple example.

\begin{Example}\label{example:polynomial codes}
We consider a PSMM problem with $V=m=p=n=T=2$. The matrices $\mathbf{A}$ and $\mathbf{B}^{(1)},\mathbf{B}^{(2)}$ are partitioned as
\begin{IEEEeqnarray}{c} 
\mathbf{A}\!\!=\!\!
\left[
  \begin{array}{@{}c@{\;}c@{}}
    \mathbf{A}_{1,1} & \mathbf{A}_{1,2} \\
    \mathbf{A}_{2,1} & \mathbf{A}_{2,2} \\
  \end{array}
\right]\!, \mathbf{B}^{(1)}\!\!=\!\!
\left[
  \begin{array}{@{}c@{\;}c@{}}
    \mathbf{B}_{1,1}^{(1)} & \mathbf{B}_{1,2}^{(1)} \\
    \mathbf{B}_{2,1}^{(1)} & \mathbf{B}_{2,2}^{(1)} \\
  \end{array}
\right]\!, 
\mathbf{B}^{(2)}\!\!=\!\!
\left[
  \begin{array}{@{}c@{\;}c@{}}
    \mathbf{B}_{1,1}^{(2)} & \mathbf{B}_{1,2}^{(2)} \\
    \mathbf{B}_{2,1}^{(2)} & \mathbf{B}_{2,2}^{(2)} \\
  \end{array}
\right]\!.\label{desired product:exam222} \IEEEeqnarraynumspace
\end{IEEEeqnarray}
Assume that the master wishes to privately compute $\mathbf{C}^{(1)}=\mathbf{A}\mathbf{B}^{(1)}$, which is given by
\begin{IEEEeqnarray}{rCl}
\mathbf{C}^{(1)}&\!=\!&
\left[
  \begin{array}{@{}c@{\;\;}c@{}}
    \mathbf{C}_{1,1}^{(1)} & \mathbf{C}_{1,2}^{(1)} \\
    \mathbf{C}_{2,1}^{(1)} & \mathbf{C}_{2,2}^{(1)} \\
  \end{array}
\right]\notag\\
&\!=\!&\left[
  \begin{array}{@{}c@{\;\;}c@{}}
    \mathbf{A}_{1,1}\mathbf{B}_{1,1}^{(1)}+\mathbf{A}_{1,2}\mathbf{B}_{2,1}^{(1)} & \mathbf{A}_{1,1}\mathbf{B}_{1,2}^{(1)}+\mathbf{A}_{1,2}\mathbf{B}_{2,2}^{(1)} \\
    \mathbf{A}_{2,1}\mathbf{B}_{1,1}^{(1)}+\mathbf{A}_{2,2}\mathbf{B}_{2,1}^{(1)} & \mathbf{A}_{2,1}\mathbf{B}_{1,2}^{(1)}+\mathbf{A}_{2,2}\mathbf{B}_{2,2}^{(1)} \\
  \end{array}
\right]. \label{desired product:exam} \IEEEeqnarraynumspace
\end{IEEEeqnarray}

Consider an SMM strategy based on polynomial codes with recovery threshold $K=17$ and the following assignment \cite{Zhu_SDMM}:
\begin{IEEEeqnarray}{rClrClrClrClrClrCl}
a_{1,1}&=&0,&\,\,\, a_{1,2}&=&1,&\,\,\, a_{2,1}&=&6,&\,\,\, a_{2,2}&=&7,&\,\,\, c_{1}&=&10,&\,\,\, c_2&=&11, \notag\\
b_{1,1}&=&1,&\,\,\, b_{1,2}&=&3,&\,\,\, b_{2,1}&=&0,&\,\,\, b_{2,2}&=&2,&\,\,\, d_{1}&=&4,&\,\,\, d_2&=&5. \notag
\end{IEEEeqnarray}
Then the encoding functions of $\mathbf{A}$ and $\mathbf{B}^{(1)}$ are in the forms of
\begin{IEEEeqnarray}{rCl}
f_{\mathbf{A}}(x)&\!=\!&\mathbf{A}_{1,1}\!\!+\!\!\mathbf{A}_{1,2}x\!\!+\!\!\mathbf{A}_{2,1}x^{6}\!\!+\!\!\mathbf{A}_{2,2}x^{7}\!\!+\!\!\mathbf{Z}_{1}^{\mathbf{A}}x^{10}\!\!+\!\!\mathbf{Z}_{2}^{\mathbf{A}}x^{11}, \IEEEeqnarraynumspace \\
h_{\mathbf{B}}(x)&\!=\!&\mathbf{B}_{2,1}^{(1)}\!\!+\!\!\mathbf{B}_{1,1}^{(1)}x\!\!+\!\!\mathbf{B}_{2,2}^{(1)}x^{2}\!\!+\!\!\mathbf{B}_{1,2}^{(1)}x^{3}\!\!+\!\!\mathbf{Z}_{1}^{\mathbf{B}}x^{4}\!\!+\!\!\mathbf{Z}_{2}^{\mathbf{B}}x^{5}, \label{example:encoding function:B}\IEEEeqnarraynumspace
\end{IEEEeqnarray}
where $\mathbf{Z}_{1}^{\mathbf{A}},\mathbf{Z}_{2}^{\mathbf{A}}$ and $\mathbf{Z}_{1}^{\mathbf{B}},\mathbf{Z}_{2}^{\mathbf{B}}$ are the matrices with corresponding dimensions and will be specified later.
The computation $\mathbf{C}^{(1)}$ can be completed by interpolating the polynomial $g(x)=\sum_{r=0}^{16}g_rx^r=f_{\mathbf{A}}(x)\cdot h_{\mathbf{B}}(x)$ because
\begin{IEEEeqnarray*}{rCl}
g_{1}&=&\mathbf{A}_{1,1}\mathbf{B}_{1,1}^{(1)}+\mathbf{A}_{1,2}\mathbf{B}_{2,1}^{(1)}=\mathbf{C}_{1,1}^{(1)},  \\ g_{3}&=&\mathbf{A}_{1,1}\mathbf{B}_{1,2}^{(1)}+\mathbf{A}_{1,2}\mathbf{B}_{2,2}^{(1)}=\mathbf{C}_{1,2}^{(1)}, \\
g_{7}&=&\mathbf{A}_{2,1}\mathbf{B}_{1,1}^{(1)}+\mathbf{A}_{2,2}\mathbf{B}_{2,1}^{(1)}=\mathbf{C}_{2,1}^{(1)}, \\ g_{9}&=&\mathbf{A}_{2,1}\mathbf{B}_{1,2}^{(1)}+\mathbf{A}_{2,2}\mathbf{B}_{2,2}^{(1)}=\mathbf{C}_{2,2}^{(1)}. \notag
\end{IEEEeqnarray*}

In PSMM, to ensure security, let $\mathbf{Z}_{1}^{\mathbf{A}},\mathbf{Z}_{2}^{\mathbf{A}}$ be independently and uniformly random matrices over $\mathbb{F}_q$. Then the master shares $\widetilde{\mathbf{A}}_i=f_{\mathbf{A}}(\alpha_i)$ with worker $i$, where $\alpha_1,\alpha_2,\ldots,\alpha_N$ are pairwise distinct non-zero elements from $\mathbb{F}_q$. Along with $\widetilde{\mathbf{A}}_i$, for the partitioning sub-matrices $\mathbf{B}_{\ell,j}^{(1)}$ and $\mathbf{B}_{\ell,j}^{(2)}$, the master also shares with the queries  $q_{\ell,j}^{(1)}(\alpha_i)$ and $q_{\ell,j}^{(2)}(\alpha_i)$, respectively, for all $\ell,j =1,2$, which are given by
\begin{IEEEeqnarray}{rCll}
q_{\ell,j}^{(1)}(\alpha_i)&=&\alpha_i^{b_{\ell,j}}+z^{(1)}_{\ell,j,1}\cdot \alpha_i^{4}+z^{(1)}_{\ell,j,2}\cdot \alpha_i^{5},\notag\\
q_{\ell,j}^{(2)}(\alpha_i)&=&z^{(2)}_{\ell,j,1}\cdot \alpha_i^{4}+z^{(2)}_{\ell,j,2}\cdot \alpha_i^{5},\notag
\end{IEEEeqnarray}
where $z^{(1)}_{\ell,j,1},z^{(1)}_{\ell,j,2}$ and $z^{(2)}_{\ell,j,1},z^{(2)}_{\ell,j,2}$ are uniformly random noises in $\mathbb{F}_q$ that protect $T=2$ colluding privacy. Upon the queries, each worker $i$ encodes $\mathbf{B}^{(1)},\mathbf{B}^{(2)}$ into
\begin{IEEEeqnarray}{rCl}
\widetilde{\mathbf{B}}_i&=&\mathbf{B}_{1,1}^{(1)} q_{1,1}^{(1)}(\alpha_i)+\mathbf{B}_{1,2}^{(1)} q_{1,2}^{(1)}(\alpha_i)+\mathbf{B}_{2,1}^{(1)} q_{2,1}^{(1)}(\alpha_i)\notag\\
&&+\mathbf{B}_{2,2}^{(1)} q_{2,2}^{(1)}(\alpha_i)+\mathbf{B}_{1,1}^{(2)} q_{1,1}^{(2)}(\alpha_i)+\mathbf{B}_{1,2}^{(2)}q_{1,2}^{(2)}(\alpha_i)\notag\\
&&+\mathbf{B}_{2,1}^{(2)} q_{2,1}^{(2)}(\alpha_i)+\mathbf{B}_{2,2}^{(2)} q_{2,2}^{(2)}(\alpha_i)\notag\\
&=&\mathbf{B}_{2,1}^{(1)}+\mathbf{B}_{1,1}^{(1)}\alpha_i+\mathbf{B}_{2,2}^{(1)}\alpha_i^{2}+\mathbf{B}_{1,2}^{(1)}\alpha_i^{3}+\mathbf{Z}_{1}^{\mathbf{B}}\alpha_i^{4}+\mathbf{Z}_{2}^{\mathbf{B}}\alpha_i^{5}\notag\\
&=&h_{\mathbf{B}}(\alpha_i),\notag
\end{IEEEeqnarray}
where we set $\mathbf{Z}_{t}^{\mathbf{B}}=\mathbf{B}_{1,1}^{(1)}z^{(1)}_{1,1,t}+\mathbf{B}_{1,2}^{(1)}z^{(1)}_{1,2,t}+\mathbf{B}_{2,1}^{(1)}z^{(1)}_{2,1,t}+\mathbf{B}_{2,2}^{(1)}z^{(1)}_{2,2,t}+
\mathbf{B}_{1,1}^{(2)}z^{(2)}_{1,1,t}+\mathbf{B}_{1,2}^{(2)}z^{(2)}_{1,2,t}+\mathbf{B}_{2,1}^{(2)}z^{(2)}_{2,1,t}+\mathbf{B}_{2,2}^{(2)}z^{(2)}_{2,2,t}$ for any $t=1,2$ that is constant for all workers and align interference to the dimension corresponding to $x^{4}$ if $t=1$ or $x^{5}$ if $t=2$, similar to \eqref{example:encoding function:B}.

Each worker $i$ computes $\widetilde{\mathbf{A}}_i\widetilde{\mathbf{B}}_i$ as a response, which can be viewed as evaluating of $g(x)=f_{\mathbf{A}}(x)\cdot h_{\mathbf{B}}(x)$ at point $x=\alpha_i$. Hence, the master can interpolate the product $g(x)$ from any $K=\deg(g(x))+1=17$ responses, and recovers the desired computation $\mathbf{A}\mathbf{B}^{(1)}$.
\end{Example}

\vspace{2mm}
Next, we formally describe the general PSMM construction based on polynomial codes.
Let the positive integers $\{a_{k,\ell},b_{\ell,j},c_t,d_t:k\in[m],\ell\in[p],j\in[n],t\in[T]\}$ satisfying C1-C2 be the parameters of the SMM strategy based on polynomial codes.
To ensure the security of its own matrix $\mathbf{A}$ \eqref{GT1:partion1}, the master employs the encoding function defined in \eqref{encoded marix:C1} to encode $\mathbf{A}$ as
\begin{IEEEeqnarray}{c}\label{PSMM:encodingfunction:A}
f_{\mathbf{A}}(x)=\sum\limits_{k=1}^{m}\sum\limits_{\ell=1}^{p}\mathbf{A}_{k,\ell}x^{a_{k,\ell}}+\sum\limits_{t=1}^{T}\mathbf{Z}_{t}^{\mathbf{A}}x^{c_t},
\end{IEEEeqnarray}
where $\mathbf{Z}_{1}^{\mathbf{A}},\ldots,\mathbf{Z}_{T}^{\mathbf{A}}$ are the random matrices over $\mathbb{F}_q$ with the same dimension as $\mathbf{A}_{k,\ell}$.

To keep the index $\theta$ private, the master generates $VTpn$ random noises $\{z^{(v)}_{\ell,j,t}:\ell\in[p],j\in[n],v\in[V],t\in[T]\}$ independently and uniformly from $\mathbb{F}_q$. Then based on the structure of the encoding function defined in \eqref{encoded marix:C2}, construct  the query polynomial $q_{\ell,j}^{(v)}(x)$ for any $\ell\in[p],j\in[n],v\in[V]$ as
\begin{IEEEeqnarray}{rCll}\label{query:poly}
q_{\ell,j}^{(v)}(x)&=&\sum\limits_{t=1}^{T}z^{(v)}_{\ell,j,t}\cdot x^{d_t}+\left\{
\begin{array}{@{}ll}
x^{b_{\ell,j}}, &\mathrm{if}\,\, v=\theta\\
0, & \mathrm{if}\,\, v\neq\theta
\end{array}
\right..
\end{IEEEeqnarray}
Here the number of the query polynomials is deliberately designed to be equal to the number of the partitioning sub-matrices in the library $\mathcal{L}^{\mathbf{B}}$.
In particular, each query polynomial $q_{\ell,j}^{(v)}(x)$ corresponds to the sub-matrix $\mathbf{B}_{\ell,j}^{(v)}$ for any $\ell\in[p],j\in[n],v\in[V]$, and will be used as encoding coefficient to encode $\mathbf{B}_{\ell,j}^{(v)}$, as shown in \eqref{encoding:B}. When $v=\theta$, the sub-matrix $\mathbf{B}_{\ell,j}^{(v)}$ is desired to be computed and the coefficient $x^{b_{\ell,j}}$ is used to encode $\mathbf{B}_{\ell,j}^{(v)}$ in the same sense as the encoding function in \eqref{encoded marix:C2}. When $v\neq\theta$, the coefficient $0$ is used to eliminate interference from the undesired sub-matrix $\mathbf{B}_{\ell,j}^{(v)}$. Moreover, the term $\sum_{t=1}^{T}z^{(v)}_{\ell,j,t}\cdot x^{d_t}$ is to provide robustness against $T$ colluding privacy, and to align interference from all the partitioning sub-matrices in $\mathcal{L}^{\mathbf{B}}$ (see \eqref{align noise}) since it has identical structure across all these partitioning sub-matrices.

Let $\alpha_1,\alpha_2,\ldots,\alpha_N$ be the pairwise distinct non-zero elements in $\mathbb{F}_q$. The master shares the evaluations of $f_{\mathbf{A}}(x)$ and $\{q_{\ell,j}^{(v)}(x):\ell\in[p],j\in[n],v\in[V]\}$ at point $x=\alpha_i$ with worker $i$, i.e.,
\begin{IEEEeqnarray}{rCl}
\widetilde{\mathbf{A}}_i&=&f_{\mathbf{A}}(\alpha_i), \label{encoding:A}\\
\mathcal{Q}_{i}^{(\theta)}&=&\{q_{\ell,j}^{(v)}(\alpha_i):\ell\in[p],j\in[n],v\in[V]\}. \label{query:PSMM:PC}
\end{IEEEeqnarray}

After receiving the query $\mathcal{Q}_{i}^{(\theta)}$, worker $i$ encodes the matrices $\mathbf{B}^{([V])}$ by taking a linear combination of the elements in $\mathcal{Q}_{i}^{(\theta)}$ and all the partitioning sub-matrices of $\mathbf{B}^{([V])}$ \eqref{GT1:partion1}, given by
\begin{IEEEeqnarray}{rCl}\label{encoding:B}
\widetilde{\mathbf{B}}_i=\sum\limits_{v=1}^{V}\sum\limits_{\ell=1}^{p}\sum\limits_{j=1}^{n}\mathbf{B}_{\ell,j}^{(v)}\cdot q_{\ell,j}^{(v)}(\alpha_i).
\end{IEEEeqnarray}

Denote the encoding function of $\mathbf{B}^{([V])}$ by
\begin{IEEEeqnarray}{rCl}
h_{\mathbf{B}}(x)&\!=\!&\sum\limits_{v=1}^{V}\sum\limits_{\ell=1}^{p}\sum\limits_{j=1}^{n}\mathbf{B}_{\ell,j}^{(v)}\cdot q_{\ell,j}^{(v)}(x) \label{encoding:A:poly}\\
&\!\overset{(a)}{=}\!&\sum\limits_{\ell=1}^{p}\sum\limits_{j=1}^{n}\mathbf{B}_{\ell,j}^{(\theta)}x^{b_{\ell,j}}\!+\!\sum\limits_{v=1}^{V}\sum\limits_{\ell=1}^{p}\sum\limits_{j=1}^{n}\mathbf{B}_{\ell,j}^{(v)}\!\cdot\!\sum\limits_{t=1}^{T}z^{(v)}_{\ell,j,t}\!\cdot\! x^{d_t}\notag\\
&\!=\!&\sum\limits_{\ell=1}^{p}\sum\limits_{j=1}^{n}\mathbf{B}_{\ell,j}^{(\theta)}x^{b_{\ell,j}}\!+\!\sum\limits_{t=1}^{T}x^{d_t}\cdot\sum\limits_{v=1}^{V}\sum\limits_{\ell=1}^{p}\sum\limits_{j=1}^{n}\mathbf{B}_{\ell,j}^{(v)} z^{(v)}_{\ell,j,t} \notag\\
&\!=\!&\sum\limits_{\ell=1}^{p}\sum\limits_{j=1}^{n}\mathbf{B}_{\ell,j}^{(\theta)}x^{b_{\ell,j}}\!+\!\sum\limits_{t=1}^{T}\mathbf{Z}_{t}^{\mathbf{B}}x^{d_t}, \label{Tans:encoding function}
\end{IEEEeqnarray}
where $(a)$ follows by \eqref{query:poly}, and
\begin{IEEEeqnarray}{c}\label{align noise}
\mathbf{Z}_{t}^{\mathbf{B}}=\sum\limits_{v=1}^{V}\sum\limits_{\ell=1}^{p}\sum\limits_{j=1}^{n}\mathbf{B}_{\ell,j}^{(v)}z^{(v)}_{\ell,j,t},\quad\forall\,t\in[T]
\end{IEEEeqnarray}
which are identical for all workers and thus can be viewed as constant terms.

We can observe from \eqref{Tans:encoding function}-\eqref{align noise} that, the encoding function $h_{\mathbf{B}}(x)$ of $\mathbf{B}^{([V])}$ efficiently separates the desired sub-matrices and the interference in the same structure as the encoding function in \eqref{encoded marix:C2}, where the desired sub-matrices $\{\mathbf{B}_{\ell,j}^{(\theta)}\}_{\ell\in[p],j\in[n]}$ appear along the $pn$ dimensions corresponding to $\{x^{b_{\ell,j}}\}_{\ell\in[p],j\in[n]}$ and the interference from all the partitioning sub-matrices of $\mathbf{B}^{([V])}$ is aligned along the $T$ dimensions corresponding to $\{x^{d_t}\}_{t\in[T]}$.
Thus, we have the fact from the SMM strategy based on polynomial codes in Section \ref{SMM:PC} that, the desired computation $\mathbf{C}^{(\theta)}=\mathbf{A}\mathbf{B}^{(\theta)}$ can be recovered from the product polynomial $g(x)=f_{\mathbf{A}}(x)\cdot h_{\mathbf{B}}(x)$.

Next, each worker $i$ computes the product
\begin{IEEEeqnarray}{c}\label{PSMM:PC:response}
\mathbf{Y}_i^{(\theta)}=\widetilde{\mathbf{A}}_i\widetilde{\mathbf{B}}_i
\end{IEEEeqnarray}
 and send it back to the master, which is equivalent to evaluating of the polynomial $g(x)=f_{\mathbf{A}}(x)\cdot h_{\mathbf{B}}(x)$ at point $x=\alpha_i$ by \eqref{encoding:A} and \eqref{encoding:B}-\eqref{encoding:A:poly}.
Thus, the master can interpolate $g(x)$ from the responses of any $K=\deg(g(x))+1=\max\{a_{k,\ell},c_t:k\in[m],\ell\in[p],t\in[T]\}+\max\{b_{\ell,j},d_t:\ell\in[p],j\in[n],t\in[T]\}+1$ workers and then recovers $\mathbf{C}^{(\theta)}=\mathbf{A}\mathbf{B}^{(\theta)}$. 

\subsection{PSMM Strategy Based on Lagrange Codes}\label{section:PSMM:LC}
Now we state the strategy that proves Theorem \ref{theorem:PSMM:LC}. We will show that, given any upper bound construction $a=(a_{r,k,\ell}),b=(b_{r,\ell,j}),c=(c_{r,k,j})$ with rank $R$ for bilinear complexity of multiplying an $m$-by-$p$ matrix and a $p$-by-$n$ matrix, the SMM strategy based on Lagrange codes can be exploited to construct a PSMM strategy with same recovery threshold. Let us start with an example to illustrate the idea.

\begin{Example}
Consider the same parameters as Example \ref{example:polynomial codes}, i.e., $V=m=p=n=T=2$. The partitions of matrices $\mathbf{A},\mathbf{B}^{(1)},\mathbf{B}^{(2)}$ and the desired computation $\mathbf{C}^{(1)}=\mathbf{A}\mathbf{B}^{(1)}$ are shown in \eqref{desired product:exam222} and \eqref{desired product:exam}, respectively.

Firstly, we use Strassen's construction \cite{Strassen} with bilinear complexity $R=7$ to encode the sub-matrices of $\mathbf{A}$ and $\mathbf{B}^{(1)},\mathbf{B}^{(2)}$ as
\begin{IEEEeqnarray*}{rClrClrCl}
\mathbf{A}_{1}\!&\!=\!&\!\mathbf{A}_{1,1}\!+\!\mathbf{A}_{2,2},&\;\; \mathbf{B}^{(1)}_{1}\!&\!\!=\!\!&\!\mathbf{B}^{(1)}_{1,1}\!+\!\mathbf{B}^{(1)}_{2,2},&\;\; \mathbf{B}^{(2)}_{1}\!&\!\!=\!\!&\!\mathbf{B}^{(2)}_{1,1}\!+\!\mathbf{B}^{(2)}_{2,2},\\
\mathbf{A}_{2}\!&\!=\!&\!\mathbf{A}_{2,1}\!+\!\mathbf{A}_{2,2},&\;\; \mathbf{B}^{(1)}_{2}\!&\!=\!&\!\mathbf{B}^{(1)}_{1,1},&\;\; \mathbf{B}^{(2)}_{2}\!&\!=\!&\!\mathbf{B}^{(2)}_{1,1},\\
\mathbf{A}_{3}\!&\!=\!&\!\mathbf{A}_{1,1},&\;\; \mathbf{B}^{(1)}_{3}\!&\!=\!&\!\mathbf{B}^{(1)}_{1,2}-\mathbf{B}^{(1)}_{2,2},&\;\; \mathbf{B}^{(2)}_{3}\!&\!=\!&\!\mathbf{B}^{(2)}_{1,2}-\mathbf{B}^{(2)}_{2,2},\\
\mathbf{A}_{4}\!&\!=\!&\!\mathbf{A}_{2,2},&\;\; \mathbf{B}^{(1)}_{4}\!&\!=\!&\!\mathbf{B}^{(1)}_{2,1}-\mathbf{B}^{(1)}_{1,1},&\;\; \mathbf{B}^{(2)}_{4}\!&\!=\!&\!\mathbf{B}^{(2)}_{2,1}-\mathbf{B}^{(2)}_{1,1},\\
\mathbf{A}_{5}\!&\!=\!&\!\mathbf{A}_{1,1}\!+\!\mathbf{A}_{1,2},&\;\; \mathbf{B}^{(1)}_{5}\!&\!=\!&\!\mathbf{B}^{(1)}_{2,2},&\;\; \mathbf{B}^{(2)}_{5}\!&\!=\!&\!\mathbf{B}^{(2)}_{2,2},\\
\mathbf{A}_{6}\!&\!=\!&\!\mathbf{A}_{2,1}-\mathbf{A}_{1,1},&\;\; \mathbf{B}^{(1)}_{6}\!&\!=\!&\!\mathbf{B}^{(1)}_{1,1}\!+\!\mathbf{B}^{(1)}_{1,2},&\;\; \mathbf{B}^{(2)}_{6}\!&\!=\!&\!\mathbf{B}^{(2)}_{1,1}\!+\!\mathbf{B}^{(2)}_{1,2},\\
\mathbf{A}_{7}\!&\!=\!&\!\mathbf{A}_{1,2}-\mathbf{A}_{2,2},&\;\; \mathbf{B}^{(1)}_{7}\!&\!=\!&\!\mathbf{B}^{(1)}_{2,1}\!+\!\mathbf{B}^{(1)}_{2,2},&\;\; \mathbf{B}^{(2)}_{7}\!&\!=\!&\!\mathbf{B}^{(2)}_{2,1}\!+\!\mathbf{B}^{(2)}_{2,2}.
\end{IEEEeqnarray*}
The desired computation $\mathbf{C}^{(1)}$ in \eqref{desired product:exam} can be recovered from the element-wise product $\{\mathbf{A}_{r}\mathbf{B}_{r}^{(1)}\}_{r\in[7]}$ by
\begin{IEEEeqnarray*}{rCl}
\mathbf{C}^{(1)}_{1,1}&=&\mathbf{A}_{1}\mathbf{B}^{(1)}_{1}+\mathbf{A}_{4}\mathbf{B}^{(1)}_{4}-\mathbf{A}_{5}\mathbf{B}^{(1)}_{5}+\mathbf{A}_{7}\mathbf{B}^{(1)}_{7},\\
\mathbf{C}^{(1)}_{1,2}&=&\mathbf{A}_{3}\mathbf{B}^{(1)}_{3}+\mathbf{A}_{5}\mathbf{B}^{(1)}_{5},\\
\mathbf{C}^{(1)}_{2,1}&=&\mathbf{A}_{2}\mathbf{B}^{(1)}_{2}+\mathbf{A}_{4}\mathbf{B}^{(1)}_{4},\\
\mathbf{C}^{(1)}_{2,2}&=&\mathbf{A}_{1}\mathbf{B}^{(1)}_{1}-\mathbf{A}_{2}\mathbf{B}^{(1)}_{2}+\mathbf{A}_{3}\mathbf{B}^{(1)}_{3}+\mathbf{A}_{6}\mathbf{B}^{(1)}_{6}.
\end{IEEEeqnarray*}

Let $\{\beta_{r},\alpha_i:r\in[9],i\in[N]\}$ be $N+9$ distinct elements from $\mathbb{F}_q$.
Further, we know from SMM based on Lagrange codes that, the element-wise product $\{\mathbf{A}_{r}\mathbf{B}_{r}^{(1)}\}_{r\in[7]}$ can be obtained by evaluating of the product polynomial $g(x)=f_{\mathbf{A}}(x)\cdot h_{\mathbf{B}}(x)$ at points $x=\beta_{1},\ldots,\beta_{7}$, where
\begin{IEEEeqnarray}{rCl}
f_{\mathbf{A}}(x)&=&\sum\limits_{r=1}^{7}\mathbf{A}_{r}\cdot\prod_{j\in[9]\backslash\{r\}}\frac{x-\beta_{j}}{\beta_{r}-\beta_{j}}\notag\\
&&+\mathbf{Z}_{8}^{\mathbf{A}}\cdot\prod_{j\in[9]\backslash\{8\}}\frac{x-\beta_{j}}{\beta_{8}-\beta_{j}}+\mathbf{Z}_{9}^{\mathbf{A}}\cdot\prod_{j\in[9]\backslash\{9\}}\frac{x-\beta_{j}}{\beta_{9}-\beta_{j}},\notag\\
h_{\mathbf{B}}(x)&=&\sum\limits_{r=1}^{7}\mathbf{B}_{r}^{(1)}\cdot\prod_{j\in[9]\backslash\{r\}}\frac{x-\beta_{j}}{\beta_{r}-\beta_{j}}\notag\\
&&+\mathbf{Z}_{8}^{\mathbf{B}}\cdot\prod_{j\in[9]\backslash\{8\}}\frac{x-\beta_{j}}{\beta_{8}-\beta_{j}}
+\mathbf{Z}_{9}^{\mathbf{B}}\cdot\prod_{j\in[9]\backslash\{9\}}\frac{x-\beta_{j}}{\beta_{9}-\beta_{j}}.\notag
\end{IEEEeqnarray}
Here $\mathbf{Z}_{8}^{\mathbf{A}},\mathbf{Z}_{9}^{\mathbf{A}}$ and $\mathbf{Z}_{8}^{\mathbf{B}},\mathbf{Z}_{9}^{\mathbf{B}}$ are the matrices with corresponding dimensions and will be specified later.

In PSMM, let $\mathbf{Z}_{8}^{\mathbf{A}},\mathbf{Z}_{9}^{\mathbf{A}}$ be random matrices to be ensured security over $\mathbb{F}_q$, and the master shares $\widetilde{\mathbf{A}}_i=f_{\mathbf{A}}(\alpha_i)$ with worker $i$. Moreover, for any $r\in[7]$, the master also shares the queries $q_r^{(1)}(\alpha_i)$ and $q_r^{(2)}(\alpha_i)$ for the sub-matrices $\mathbf{B}_{r}^{(1)}$ and $\mathbf{B}_{r}^{(2)}$, respectively, which are given by
\begin{IEEEeqnarray*}{rCl}
q_r^{(1)}(\alpha_i)
&\!=\!&\prod\limits_{j\in[9]\backslash\{r\}}\frac{\alpha_i-\beta_{j}}{\beta_{r}-\beta_{j}}\notag\\
&&+z_{r,8}^{(1)}\!\cdot\!\prod_{j\in[9]\backslash\{8\}}\frac{\alpha_i-\beta_{j}}{\beta_{8}-\beta_{j}}
\!+\!z_{r,9}^{(1)}\!\cdot\!\prod_{j\in[9]\backslash\{9\}}\frac{\alpha_i-\beta_{j}}{\beta_{9}-\beta_{j}}, \\
q_r^{(2)}(\alpha_i)
&=&z_{r,8}^{(2)}\cdot\prod_{j\in[9]\backslash\{8\}}\frac{\alpha_i-\beta_{j}}{\beta_{8}-\beta_{j}}
+z_{r,9}^{(2)}\cdot\prod_{j\in[9]\backslash\{9\}}\frac{\alpha_i-\beta_{j}}{\beta_{9}-\beta_{j}},
\end{IEEEeqnarray*}
where $z_{r,8}^{(1)},z_{r,9}^{(1)}$ and $z_{r,8}^{(2)},z_{r,9}^{(2)}$ are random noises from $\mathbb{F}_q$.
Then worker $i$ encodes the library into
\begin{IEEEeqnarray*}{rCl}
\widetilde{\mathbf{B}}_i&=&\sum\limits_{r=1}^{7}\mathbf{B}_{r}^{(1)} q_r^{(1)}(\alpha_i)+\sum\limits_{r=1}^{7}\mathbf{B}_{r}^{(2)} q_r^{(2)}(\alpha_i)\\
&=&\sum\limits_{r=1}^{7}\mathbf{B}_{r}^{(1)}\cdot\prod_{j\in[9]\backslash\{r\}}\frac{\alpha_i-\beta_{j}}{\beta_{r}-\beta_{j}}\\
&&+\mathbf{Z}_{8}^{\mathbf{B}}\cdot\prod_{j\in[9]\backslash\{8\}}\frac{\alpha_i-\beta_{j}}{\beta_{8}-\beta_{j}}
+\mathbf{Z}_{9}^{\mathbf{B}}\cdot\prod_{j\in[9]\backslash\{9\}}\frac{\alpha_i-\beta_{j}}{\beta_{9}-\beta_{j}},
\end{IEEEeqnarray*}
where we set $\mathbf{Z}_{k}^{\mathbf{B}}=\sum_{r=1}^{7}\mathbf{B}_{r}^{(1)} z_{r,k}^{(1)}+\sum_{r=1}^{7}\mathbf{B}_{r}^{(2)} z_{r,k}^{(2)}$ for any $k=8,9$.
Next, worker $i$ computes $\widetilde{\mathbf{A}}_i\widetilde{\mathbf{B}}_i$ as a response, which is the evaluation of $g(x)=f_{\mathbf{A}}(x)\cdot h_{\mathbf{B}}(x)$ at point $x=\alpha_i$. Hence, the master can interpolate the product $g(x)$ from any $K=\deg(g(x))+1=17$ responses, and then recovers $\mathbf{C}^{(1)}=\mathbf{A}\mathbf{B}^{(1)}$.
\end{Example}

The general construction is described as follows. Similar to \eqref{bilinear:constru:1} and \eqref{bilinear:constru:2}, the master and each worker converts the matrices $\mathbf{A}$ and $\mathbf{B}^{(v)}$ \eqref{GT1:partion1} into a batch of sub-matrices of length $R$, respectively, as shown below.
\begin{IEEEeqnarray}{rCl}
\mathbf{A}_{r}&=&\sum\limits_{k=1}^{m}\sum\limits_{\ell=1}^{p}a_{r,k,\ell}\mathbf{A}_{k,\ell}, \quad\forall\, r\in[R],\label{BC:AA}\\
\mathbf{B}_{r}^{(v)}&=&\sum\limits_{\ell=1}^{p}\sum\limits_{j=1}^{n}b_{r,\ell,j}\mathbf{B}_{\ell,j}^{(v)}, \quad\forall\, r\in[R],v\in[V].\label{BC:B}
\end{IEEEeqnarray}
By \eqref{lag:recover}, the master can recover the desired computation $\mathbf{A}\mathbf{B}^{(\theta)}$ if it is able to obtain the element-wise product $\{\mathbf{A}_{r}\mathbf{B}_{r}^{(\theta)}:r\in[R]\}$ of the two batches of sub-matrices $(\mathbf{A}_{1},\ldots,\mathbf{A}_{R})$ and $(\mathbf{B}_{1}^{(\theta)},\ldots,\mathbf{B}_{R}^{(\theta)})$.

Let $\{\beta_{r},\alpha_i:r\in[R+T],i\in[N]\}$ be $R+T+N$ pairwise distinct elements from $\mathbb{F}_q$.
To complete the computation, the master employs the encoding function defined in \eqref{encoding funcion:lagrange:A} to encode the batch of sub-matrices $(\mathbf{A}_{1},\ldots,\mathbf{A}_{R})$ as
\begin{IEEEeqnarray}{l}
f_{\mathbf{A}}(x)=\sum\limits_{r=1}^{R}\mathbf{A}_{r}\cdot\prod_{j\in[R+T]\backslash\{r\}}\frac{x-\beta_{j}}{\beta_{r}-\beta_{j}}\notag\\
\quad\quad\quad\quad\quad\quad\quad+\sum\limits_{k=R+1}^{R+T}\mathbf{Z}_{k}^{\mathbf{A}}\cdot\prod_{j\in[R+T]\backslash\{k\}}\frac{x-\beta_{j}}{\beta_{k}-\beta_{j}}, \label{PSMM:PL:encdoingAA}\IEEEeqnarraynumspace
\end{IEEEeqnarray}
where $\mathbf{Z}_{R+1}^{\mathbf{A}},\ldots,\mathbf{Z}_{R+T}^{\mathbf{A}}$ are random matrices over $\mathbb{F}_q$ with the same dimension as $\mathbf{A}_{r}$.

To keep the index $\theta$ private, let $\{z_{r,k}^{(v)}:k\in[R+1:R+T],r\in[R],v\in[V]\}$ be $TRV$ independently and uniformly random noises from $\mathbb{F}_q$. Then, given any $v\in[V],r\in[R]$, based on the structure of the encoding function defined in \eqref{encoding funcion:lagrange:B}, the master constructs the query polynomial $q_r^{(v)}(x)$ as
\begin{IEEEeqnarray}{l}
q_r^{(v)}(x)
=\sum\limits_{k=R+1}^{R+T}z_{r,k}^{(v)}\cdot\prod_{j\in[R+T]\backslash\{k\}}\frac{x-\beta_{j}}{\beta_{k}-\beta_{j}}\notag\\
\quad\quad\quad\quad\quad\quad\quad\quad+\left\{
\begin{array}{@{}l@{\;\;}l}
\prod\limits_{j\in[R+T]\backslash\{r\}}\frac{x-\beta_{j}}{\beta_{r}-\beta_{j}},
&\mathrm{if}\,\, v=\theta \\
0, &\mathrm{if}\,\, v\neq\theta
\end{array}
\right.. \label{PSMM:query}\IEEEeqnarraynumspace
\end{IEEEeqnarray}

Then the master evaluates $f_{\mathbf{A}}(x)$ and $\{q_r^{(v)}(x):r\in[R],v\in[V]\}$ at point $x=\alpha_i$, and sends them to worker $i$, i.e.,
\begin{IEEEeqnarray}{rCl}
\widetilde{\mathbf{A}}_i&=&f_{\mathbf{A}}(\alpha_i),\label{SPMM:encoding:A}\\
\mathcal{Q}_{i}^{(\theta)}&=&\{q_r^{(v)}(\alpha_i):r\in[R],v\in[V]\}. \label{SPMM:encoding:AAAA}
\end{IEEEeqnarray}

Upon the received query, worker $i$ encodes the matrices $\mathbf{B}^{([V])}$ into
\begin{IEEEeqnarray}{c}\label{PSMM:PC:encodingBB}
\widetilde{\mathbf{B}}_i=\sum\limits_{v=1}^{V}\sum\limits_{r=1}^{R}\mathbf{B}_{r}^{(v)}\cdot q_r^{(v)}(\alpha_i).
\end{IEEEeqnarray}

By \eqref{PSMM:query}, we can denote the encoding function $h_{\mathbf{B}}(x)$ of matrices $\mathbf{B}^{([V])}$ by
\begin{IEEEeqnarray}{rCl}
h_{\mathbf{B}}(x)&=&\sum\limits_{v=1}^{V}\sum\limits_{r=1}^{R}\mathbf{B}_{r}^{(v)}\cdot q_r^{(v)}(x) \notag\\
&=&\sum\limits_{r=1}^{R}\mathbf{B}_{r}^{(\theta)}\cdot \prod\limits_{j\in[R+T]\backslash\{r\}}\frac{x-\beta_{j}}{\beta_{r}-\beta_{j}}\notag\\
&&\quad\quad\quad+\sum\limits_{k=R+1}^{R+T}\mathbf{Z}_{k}^{\mathbf{B}}\cdot\prod_{j\in[R+T]\backslash\{k\}} \frac{x-\beta_{j}}{\beta_{k}-\beta_{j}},\label{encoding funtion:matrixB} \IEEEeqnarraynumspace
\end{IEEEeqnarray}
where
\begin{IEEEeqnarray}{c}
\mathbf{Z}_{k}^{\mathbf{B}}=\sum\limits_{v=1}^{V}\sum\limits_{r=1}^{R}\mathbf{B}_{r}^{(v)}\cdot z_{r,k}^{(v)}.\notag
\end{IEEEeqnarray}
Note from \eqref{encoding funtion:matrixB} that, the function $h_{\mathbf{B}}(x)$ has the identical structure as $h_L(x)$ in \eqref{encoding funcion:lagrange:B}, which aligns the batch of desired sub-matrices $\{\mathbf{B}_{r}^{(\theta)}\}_{r\in[R]}$ along the $R$ dimensions corresponding to $\{\prod_{j\in[R+T]\backslash\{r\}}\frac{x-\beta_{j}}{\beta_{r}-\beta_{j}}\}_{r\in[R]}$ and the interference from the matrices $\mathbf{B}^{([V])}$ is aligned along the $T$ dimensions corresponding to $\{\prod_{j\in[R+T]\backslash\{k\}}\frac{x-\beta_{j}}{\beta_{k}-\beta_{j}}\}_{k\in[R+1:R+T]}$.

Then worker $i$ computes  $\mathbf{Y}_i^{(\theta)}=\widetilde{\mathbf{A}}_i\widetilde{\mathbf{B}}_i$, which is the evaluation of the product polynomial $g(x)=f_{\mathbf{A}}(x)\cdot h_{\mathbf{B}}(x)$ at point $x=\alpha_i$.
Thus, the master can interpolate the polynomial $g(x)$ from the responses of any $K=\deg(g(x))+1=2R+2T-1$ workers, and then evaluates $g(x)$ at points $x=\beta_1,\ldots,\beta_R$ to obtain the desired element-wise product $\{\mathbf{A}_{r}\mathbf{B}_{r}^{(\theta)}:r\in[R]\}$.

\subsection{Security, Privacy, Communication Cost and Computation Complexity for PSMM}\label{proof:strategy}
In this subsection, we prove the security and privacy of the two PSMM strategies above, analyse their communication cost and computation complexities, and compare their performance.

\subsubsection{Security and Privacy}\label{PSMM:Privacy22}
In the proposed PSMM strategy based on polynomial codes, the encoding matrices $\widetilde{\mathbf{A}}_i$ \eqref{encoding:A} and the queries $\mathcal{Q}_{i}^{(\theta)}$ \eqref{query:PSMM:PC} sent to workers are generated by evaluating the encoding polynomial $f_{\mathbf{A}}(x)$ of matrix $\mathbf{A}$ and the query polynomials $\{q_{\ell,j}^{(v)}(x)\}_{\ell\in[p],j\in[n],v\in[V]}$ at distinct points, respectively. Here the encoding polynomial $f_{\mathbf{A}}(x)$ \eqref{PSMM:encodingfunction:A} (resp. each of query polynomial $q_{\ell,j}^{(v)}(x)$ \eqref{query:poly}) is constructed by employing $T$ independent and uniform random noises to mask the confidential matrix $\mathbf{A}$ (resp. the interested index $\theta$), which ensures that the data sent to any $T$ workers are secure (resp. private). The security and privacy of the PSMM strategy based on Lagrange codes follows from similar argument. Their formal proofs are presented in Appendix-A.

\subsubsection{Communication Cost}
In the two PSMM strategies, the master sends an encoding sub-matrix with the same dimension of $\frac{\lambda}{m}\times\frac{\omega}{p}$ to each worker by \eqref{encoding:A} and \eqref{SPMM:encoding:A}, and downloads a matrix with the same dimension of $\frac{\lambda}{m}\times\frac{\gamma}{n}$ from each of responsive workers.
Thus, by \eqref{upload and download}, the two strategies achieve the same upload cost $P_u=\frac{N\times\frac{\lambda}{m}\times\frac{\omega}{p}}{\lambda\times\omega}=\frac{N}{mp}$ and the download cost  $P_d=\frac{K\times\frac{\lambda}{m}\times\frac{\gamma}{n}}{\lambda\times\gamma}=\frac{K}{mn}$, where $K=\max\{a_{k,\ell},c_t:k\in[m],\ell\in[p],t\in[T]\}+\max\{b_{\ell,j},d_t:\ell\in[p],j\in[n],t\in[T]\}+1$ and $K=2R+2T-1$ for the PSMM strategies based on polynomial codes and Lagrange codes, respectively.

\subsubsection{Computation Complexity} In the PSMM strategy based on polynomial codes, the encoding process for matrix $\mathbf{A}$ can be viewed as evaluating a polynomial of degree less than $N$ at $N$ points for $\frac{\lambda\omega}{mp}$ times by \eqref{PSMM:encodingfunction:A} and \eqref{encoding:A}, and decoding requires interpolating a $(K-1)$-th degree polynomial for $\frac{\lambda\gamma}{mn}$ times, where $K=\max\{a_{k,\ell},c_t:k\in[m],\ell\in[p],t\in[T]\}+\max\{b_{\ell,j},d_t:\ell\in[p],j\in[n],t\in[T]\}+1$.  It is well known \cite{Von} that the evaluation of a $k$-th degree polynomial at $k+1$ arbitrary points can be done in ${{O}}(k(\log k)^2\log\log k)$ arithmetic operations, and consequently, its dual problem, interpolation of a $k$-th degree polynomial from $k+1$ arbitrary points can be performed in the same arithmetic operations ${{O}}(k(\log k)^2\log\log k)$.
Thus, encoding and decoding achieve the complexities $O(\frac{\lambda\omega N(\log N)^2\log\log N}{mp})$ and ${{O}}(\frac{\lambda\gamma K(\log K)^2\log\log K}{mn})$, respectively.
The complexity at each worker includes generating a linear combination of $Vpn$ sub-matrices with dimension $\frac{\omega\gamma}{np}$ \eqref{encoding:B}, and multiplying  two coded sub-matrices with sizes $\frac{\lambda}{m}\times\frac{\omega}{p}$ and $\frac{\omega}{p}\times\frac{\gamma}{n}$, which requires a complexity of ${O}(V\omega\gamma+\frac{\lambda\omega\gamma}{mpn})$ at most.

In the PSMM strategy based on Lagrange codes, the encoding process for matrix $\mathbf{A}$ includes generating a bath of sub-matrices by taking a linear combination of $mp$ partitioning sub-matrices with dimension $\frac{\lambda\omega}{mp}$ for $R$ times \eqref{BC:AA}, and encoding the batch of sub-matrices by evaluating a polynomial of degree $R+T-1<N$ at $N$ points for $\frac{\lambda\omega}{mp}$ times by \eqref{PSMM:PL:encdoingAA} and \eqref{SPMM:encoding:A}, which achieves the complexity ${{O}}(R\lambda\omega+\frac{\lambda\omega N(\log N)^2\log\log N}{mp})$.
The complexity at each worker consists of generating another bath of sub-matrices by taking a linear combination of $pn$ partitioning sub-matrices with dimension $\frac{\omega\gamma}{np}$ for $VR$ times \eqref{BC:B}, computing a linear combination of $VR$ sub-matrices with dimension $\frac{\omega\gamma}{np}$ \eqref{PSMM:PC:encodingBB}, and multiplying  two coded sub-matrices with sizes $\frac{\lambda}{m}\times\frac{\omega}{p}$ and $\frac{\omega}{p}\times\frac{\gamma}{n}$, which requires a complexity of ${O}(VR\omega\gamma+\frac{\lambda\omega\gamma}{mpn})$.
Decoding requires interpolating of a $(K-1)$-th degree polynomial for $\frac{\lambda\gamma}{mn}$ times, then evaluating the polynomial at $R<K$ points for $\frac{\lambda\gamma}{mn}$ times, and finally computing the linear combinations of $R$ sub-matrices with dimension $\frac{\lambda\gamma}{mn}$ for $mn$ times \eqref{lag:recover}, which achieves the complexity $O(\frac{\lambda\gamma K(\log K)^2\log\log K}{mn}+R\lambda\gamma)$, where $K=2R+2T-1$.

\subsubsection{Performance Comparison Between the Proposed PSMM Strategies}\label{com:PSMM:2}
We summarize the performance of the two proposed PSMM strategies in Table \ref{tab:PSMM}.
Recall from Remark \ref{PSMM:remark} that which one of the two PSMM strategies achieves a smaller recovery threshold depends on the value of the security parameter $T$.  We observe from Table \ref{tab:PSMM} that, regardless of the value of $T$, the PSMM strategy based on polynomial codes outperforms the strategy based on Lagrange codes in terms of encoding complexity and worker computation complexity, with the upload cost being identical. When $K_P\leq K_L$ for some values of the parameter $T$, the PSMM strategy based on polynomial codes achieves a smaller recovery threshold, download cost, and decoding complexity, and otherwise (i.e., when $K_P>K_L$) the one based on Lagrange codes achieves a smaller recovery threshold and download cost. In the former case of $K_P\leq K_L$, the strategy based on polynomial codes is preferable to facilitate a faster execution of PSMM. However, in the latter case of $K_P>K_L$, one should choose which strategy to implement according to the system resources including the computation and communication capabilities of the master and the workers.

\begin{table*}[htbp]
\centering
\small
\caption{Performance of the proposed PSMM strategies.}\label{tab:PSMM}
\begin{tabular}{|@{\;}c@{\;}|c@{\;}|c@{\;}|}
  \hline
  & PSMM  based on Polynomial codes & PSMM  based on Lagrange codes  \\ \hline
\multirow{2}{*}{Recovery Threshold} & $K_P=\max\{a_{k,\ell},c_t:k\in[m],\ell\in[p],t\in[T]\}$ & \multirow{2}{*}{$K_L=2R+2T-1$} \\
& $+\max\{b_{\ell,j},d_t:\ell\in[p],j\in[n],t\in[T]\}+1$ &  \\ \hline
  Upload and Download $(P_u,P_d)$ & $\big(\frac{N}{mp},\frac{K_P}{mn}\big)$ &  $\big(\frac{N}{mp},\frac{K_L}{mn}\big)$  \\ \hline
  Encoding Complexity ${C}_{\mathbf{A}}$ & $O(\frac{\lambda\omega N(\log N)^2\log\log N}{mp})$  & ${{O}}(R\lambda\omega+\frac{\lambda\omega N(\log N)^2\log\log N}{mp})$ \\ \hline
  Worker Computation ${C}_{w}$ & ${O}(V\omega\gamma+\frac{\lambda\omega\gamma}{mpn})$ & ${O}(VR\omega\gamma+\frac{\lambda\omega\gamma}{mpn})$  \\ \hline
  Decoding Complexity ${C}_d$ & ${{O}}(\frac{\lambda\gamma K_P(\log K_P)^2\log\log K_P}{mn})$ & $O(\frac{\lambda\gamma K_L(\log K_L)^2\log\log K_L}{mn}+R\lambda\gamma)$  \\ \hline
  \end{tabular}
  \begin{tablenotes}
       \footnotesize
       \item[]\quad\quad\;\; Here, $K_P$ can take the value of $\min\{(m+1)(np+T)-1,(n+1)(mp+T)-1,2mpn+2T-1\}$ by \\
       \item[]\quad\quad\;\; Corollary \ref{cor:PSMM-poly}, and $R$ denotes any upper construction of bilinear complexity.
\end{tablenotes}
\end{table*}

\section{Computation Strategies for Fully Private Matrix Multiplication}\label{section:FPMM}
In this section, we present the FPMM strategies based on polynomial codes and Lagrange codes for proving Theorems \ref{theorem:FPMM:1} and \ref{theorem:FPMM:2}, respectively, and then analyse their privacy, communication cost and computation complexities.

As illustrated in Fig. \ref{FPMM}, the goal of the master in FPMM is to compute the product $\mathbf{A}^{(\theta_1)}\mathbf{B}^{(\theta_2)}$ for any $\theta_1\in[U]$ and $\theta_2\in[V]$, while keeping the indices of the desired product $\theta_1$ and $\theta_2$ private from any $T$ colluding workers.
Consider arbitrary partitioning parameters $m,p,n$, the matrices $\mathbf{A}^{(u)}$ and $\mathbf{B}^{(v)}$ are divided into $m\times p$ and $p\times n$ equal-size sub-matrices, respectively, i.e., for all $u\in[U],v\in[V]$,
\begin{IEEEeqnarray}{c}\label{FPMM:matrix partition}
\mathbf{A}^{(u)}\!=\!
\left[
  \begin{array}{@{}c@{\;\;}c@{\;\;}c@{}}
    \mathbf{A}_{1,1}^{(u)}  & \ldots & \mathbf{A}_{1,p}^{(u)} \\
    \vdots  & \ddots & \vdots \\
    \mathbf{A}_{m,1}^{(u)}  & \ldots & \mathbf{A}_{m,p}^{(u)} \\
  \end{array}
\right],\,\,
\mathbf{B}^{(v)}\!=\!
\left[
  \begin{array}{@{}c@{\;\;}c@{\;\;}c@{}}
    \mathbf{B}_{1,1}^{(v)} &  \ldots & \mathbf{B}_{1,n}^{(v)} \\
    \vdots & \ddots & \vdots \\
    \mathbf{B}_{p,1}^{(v)}  & \ldots & \mathbf{B}_{p,n}^{(v)} \\
  \end{array}
\right], \IEEEeqnarraynumspace
\end{IEEEeqnarray}
where $\mathbf{A}_{k,\ell}^{(u)}\in\mathbb{F}_q^{\frac{\lambda}{m}\times\frac{\omega}{p}}$ for any $k\in[m],\ell\in[p]$, and $\mathbf{B}_{\ell,j}^{(v)}\in\mathbb{F}_q^{\frac{\omega}{p}\times\frac{\gamma}{n}}$ for any $\ell\in[p],j\in[n]$. Thus, the desired product $\mathbf{C}^{(\theta_1,\theta_2)}=\mathbf{A}^{(\theta_1)}\mathbf{B}^{(\theta_2)}$ is given by
\begin{IEEEeqnarray}{c}\notag
\mathbf{C}^{(\theta_1,\theta_2)}=\mathbf{A}^{(\theta_1)}\mathbf{B}^{(\theta_2)}=
\left[
  \begin{array}{@{}ccc@{}}
    \mathbf{C}_{1,1}^{(\theta_1,\theta_2)} & \ldots & \mathbf{C}_{1,n}^{(\theta_1,\theta_2)} \\
    \vdots & \ddots & \vdots \\
    \mathbf{C}_{m,1}^{(\theta_1,\theta_2)} & \ldots & \mathbf{C}_{m,n}^{(\theta_1,\theta_2)} \\
  \end{array}
\right]
\end{IEEEeqnarray}
with $\mathbf{C}_{k,j}^{(\theta_1,\theta_2)}=\sum_{\ell=1}^{p}\mathbf{A}_{k,\ell}^{(\theta_1)}\mathbf{B}_{\ell,j}^{(\theta_2)}$ for any $k\in[m],j\in[n]$.

\subsection{FPMM Strategy Based on Polynomial Codes}\label{proof:theorem:FPMM}
We start with proving Theorem \ref{theorem:FPMM:1}. Similar to PSMM strategy based on polynomial codes, we show that any SMM strategy based on polynomial codes can be exploited to construct the FPMM strategy with same recovery threshold.

Let the positive integers $\{a_{k,\ell},b_{\ell,j},c_t,d_t:k\in[m],\ell\in[p],j\in[n],t\in[T]\}$ satisfying C1-C2 be the parameters of the SMM strategy based on polynomial codes.
For any $t\in[T]$, let $\mathbf{Z}_{t}^{\mathbf{A}}$ and $\mathbf{Z}_{t}^{\mathbf{B}}$ are arbitrary matrices over $\mathbb{F}_q$ with the same dimensions as $\mathbf{A}_{k,\ell}^{(\theta_1)}$ and $\mathbf{B}_{\ell,j}^{(\theta_2)}$ respectively, and their forms will be specified  later.
Note from Section \ref{SMM:PC} that, the desired computation $\mathbf{C}^{(\theta_1,\theta_2)}$ can be completed if one recovers the product polynomial $g(x)=f_{\mathbf{A}}(x)\cdot h_{\mathbf{B}}(x)$, where
\begin{IEEEeqnarray}{rCl}
f_{\mathbf{A}}(x)&=&\sum\limits_{k=1}^{m}\sum\limits_{\ell=1}^{p}\mathbf{A}_{k,\ell}^{(\theta_1)}x^{a_{k,\ell}}+\sum\limits_{t=1}^{T}\mathbf{Z}_{t}^{\mathbf{A}}x^{c_t}, \label{FPMM:encodingA}\\
h_{\mathbf{B}}(x)&=&\sum\limits_{\ell=1}^{p}\sum\limits_{j=1}^{n}\mathbf{B}_{\ell,j}^{(\theta_2)}x^{b_{\ell,j}}+\sum\limits_{t=1}^{T}\mathbf{Z}_{t}^{\mathbf{B}}x^{d_t}. \label{FPMM:encodingB}
\end{IEEEeqnarray}

To keep the index $\theta_1$ private, let $\{\tilde{z}^{(u)}_{k,\ell,t}:t\in[T],k\in[m],\ell\in[p],u\in[U]\}$ be $UTmp$ random noises chosen independently and uniformly from $\mathbb{F}_q$. Then for each partitioning sub-matrix $\mathbf{A}_{k,\ell}^{(u)}$ in $\mathbf{A}^{([U])}$  for any $k\in[m],\ell\in[p],u\in[U]$, the master constructs the query polynomial $\rho_{k,\ell}^{(u)}(x)$ based on the structure of the encoding function in \eqref{FPMM:encodingA}, which is given by
\begin{IEEEeqnarray}{c}\label{query:poly2}
\rho_{k,\ell}^{(u)}(x)=\sum\limits_{t=1}^{T}\tilde{z}^{(u)}_{k,\ell,t}\cdot x^{c_t}+\left\{
\begin{array}{@{}ll}
x^{a_{k,\ell}}, &\mathrm{if}\,\, u=\theta_1\\
0, & \mathrm{if}\,\, u\neq\theta_1
\end{array}
\right..
\end{IEEEeqnarray}
Similar to \eqref{query:poly}, based on the structure of the encoding function in \eqref{FPMM:encodingB}, the master also constructs the query polynomial $q_{\ell,j}^{(m)}(x)$ for each partitioning sub-matrix $\mathbf{B}_{\ell,j}^{(v)}$ in $\mathbf{B}^{([V])}$ for any $\ell\in[p],j\in[n],v\in[V]$, given by
\begin{IEEEeqnarray}{rCll}\label{query:poly2234}
q_{\ell,j}^{(v)}(x)&=&\sum\limits_{t=1}^{T}z^{(v)}_{\ell,j,t}\cdot x^{d_t}+\left\{
\begin{array}{@{}ll}
x^{b_{\ell,j}}, &\mathrm{if}\,\, v=\theta_2\\
0, & \mathrm{if}\,\, v\neq\theta_2
\end{array}
\right.,
\end{IEEEeqnarray}
where $z^{(v)}_{\ell,j,t}$ is uniformly random noise from $\mathbb{F}_q$.

Let $\alpha_1,\alpha_2,\ldots,\alpha_N$ be the pairwise distinct non-zero elements in $\mathbb{F}_q$. The master shares the following evaluations with worker $i$:
\begin{IEEEeqnarray}{rCl}
\mathcal{Q}_{i}^{(\theta_1)}&=&\{\rho_{k,\ell}^{(u)}(\alpha_i):k\in[m],\ell\in[p],u\in[U]\},\label{FPMM:queryA}\\
\mathcal{Q}_{i}^{(\theta_2)}&=&\{q_{\ell,j}^{(v)}(\alpha_i):\ell\in[p],j\in[n],v\in[V]\}. \label{FPMM:queryB}
\end{IEEEeqnarray}

After that, worker $i$ encodes the matrices $\mathbf{A}^{([U])}$ and $\mathbf{B}^{([V])}$ \eqref{FPMM:matrix partition} into
\begin{IEEEeqnarray}{rCl}
\widetilde{\mathbf{A}}_i&=&\sum\limits_{u=1}^{U}\sum\limits_{k=1}^{m}\sum\limits_{\ell=1}^{p}\mathbf{A}_{k,\ell}^{(u)}\cdot \rho_{k,\ell}^{(u)}(\alpha_i),\label{FPMM:answerA}\\
\widetilde{\mathbf{B}}_i&=&\sum\limits_{v=1}^{V}\sum\limits_{\ell=1}^{p}\sum\limits_{j=1}^{n}\mathbf{B}_{\ell,j}^{(v)}\cdot q_{\ell,j}^{(v)}(\alpha_i). \label{FPMM:answerB}
\end{IEEEeqnarray}

By \eqref{FPMM:encodingA}--\eqref{query:poly2234}, we have
\begin{IEEEeqnarray}{l}
\sum\limits_{u=1}^{U}\sum\limits_{k=1}^{m}\sum\limits_{\ell=1}^{p}\mathbf{A}_{k,\ell}^{(u)}\cdot \rho_{k,\ell}^{(u)}(x)\notag\\
\quad\quad\quad=\sum\limits_{k=1}^{m}\sum\limits_{\ell=1}^{p}\mathbf{A}_{k,\ell}^{(\theta_1)}x^{a_{k,\ell}}+\sum\limits_{t=1}^{T}\mathbf{Z}_{t}^{\mathbf{A}}x^{c_t}=f_{\mathbf{A}}(x),\label{FPMM:tansformationA} \IEEEeqnarraynumspace \\
\sum\limits_{v=1}^{V}\sum\limits_{\ell=1}^{p}\sum\limits_{j=1}^{n}\mathbf{B}_{\ell,j}^{(v)}\cdot q_{\ell,j}^{(v)}(x)\notag\\
\quad\quad\quad=\sum\limits_{\ell=1}^{p}\sum\limits_{j=1}^{n}\mathbf{B}_{\ell,j}^{(\theta)}x^{b_{\ell,j}}+\sum\limits_{t=1}^{T}\mathbf{Z}_{t}^{\mathbf{B}}x^{d_t}=h_{\mathbf{B}}(x),\label{FPMM:tansformationB} \IEEEeqnarraynumspace
\end{IEEEeqnarray}
where we set
\begin{IEEEeqnarray}{rCll}
\mathbf{Z}_{t}^{\mathbf{A}}&=&\sum\limits_{u=1}^{U}\sum\limits_{k=1}^{m}\sum\limits_{\ell=1}^{p}\mathbf{A}_{k,\ell}^{(u)}\tilde{z}^{(u)}_{k,\ell,t},\quad&\forall\,t\in[T],\notag\\
\mathbf{Z}_{t}^{\mathbf{B}}&=&\sum\limits_{v=1}^{V}\sum\limits_{\ell=1}^{p}\sum\limits_{j=1}^{n}\mathbf{B}_{\ell,j}^{(v)}z^{(v)}_{\ell,j,t},\quad&\forall\,t\in[T],\notag
\end{IEEEeqnarray}
which are independent of workers and thus can be viewed as constant terms.

Then each worker $i\in[N]$ computes the product $\mathbf{Y}_i^{(\theta_1,\theta_2)}=\widetilde{\mathbf{A}}_i\widetilde{\mathbf{B}}_i$ as a response for the master, which is equivalent to evaluating the product polynomial $g(x)=f_{\mathbf{A}}(x)\cdot h_{\mathbf{B}}(x)$ at point $x=\alpha_i$ by \eqref{FPMM:answerA}--\eqref{FPMM:tansformationB}.
Thus, the master can interpolate the product $g(x)$ from the responses of any $K=\max\{a_{k,\ell},c_t:k\in[m],\ell\in[p],t\in[T]\}+\max\{b_{\ell,j},d_t:\ell\in[p],j\in[n],t\in[T]\}+1$ workers and then recovers the desired computation $\mathbf{C}^{(\theta_1,\theta_2)}$.

\subsection{FPMM Strategy Based on Lagrange Codes}\label{proof:theorem:FPMM2}
We now present the FPMM strategy based on Lagrange codes for proving Theorem  \ref{theorem:FPMM:2}.
Let $a=(a_{r,k,\ell}),b=(b_{r,\ell,j}),c=(c_{r,k,j})$ be any upper bound construction with rank $R$ for bilinear complexity. Each worker first converts the matrices $\mathbf{A}^{(u)}$ and $\mathbf{B}^{(v)}$ \eqref{FPMM:matrix partition} into a batch of sub-matrices of length $R$, respectively:
\begin{IEEEeqnarray}{rCll}
\mathbf{A}_{r}^{(u)}&=&\sum\limits_{k=1}^{m}\sum\limits_{\ell=1}^{p}a_{r,k,\ell}\mathbf{A}_{k,\ell}^{(u)}, &\quad\forall\, r\in[R],u\in[U],\label{FPMM:LC:batchA}\\
\mathbf{B}_{r}^{(v)}&=&\sum\limits_{\ell=1}^{p}\sum\limits_{j=1}^{n}b_{r,\ell,j}\mathbf{B}_{\ell,j}^{(v)}, &\quad\forall\, r\in[R],v\in[V].\label{FPMM:LC:batchB}
\end{IEEEeqnarray}

Let $\{\beta_{r},\alpha_i:r\in[R+T],i\in[N]\}$ be $R+T+N$ distinct elements from $\mathbb{F}_q$. Denote the polynomial functions of $\mathbf{A}^{(\theta_1)}$ and $\mathbf{B}^{(\theta_2)}$ by
\begin{IEEEeqnarray}{rCl}
f_{\mathbf{A}}(x)&=&\sum\limits_{r=1}^{R}\mathbf{A}_{r}^{(\theta_1)}\cdot\prod_{j\in[R+T]\backslash\{r\}}\frac{x-\beta_{j}}{\beta_{r}-\beta_{j}}\notag\\
&&\quad\quad\quad+\sum\limits_{k=R+1}^{R+T}\mathbf{Z}_{k}^{\mathbf{A}}\cdot\prod_{j\in[R+T]\backslash\{k\}}\frac{x-\beta_{j}}{\beta_{k}-\beta_{j}},\label{FPMM:LC:encodingA}\IEEEeqnarraynumspace\\
h_{\mathbf{B}}(x)&=&\sum\limits_{r=1}^{R}\mathbf{B}_{r}^{(\theta_2)}\cdot\prod_{j\in[R+T]\backslash\{r\}}\frac{x-\beta_{j}}{\beta_{r}-\beta_{j}}\notag\\
&&\quad\quad\quad+\sum\limits_{k=R+1}^{R+T}\mathbf{Z}_{k}^{\mathbf{B}}\cdot\prod_{j\in[R+T]\backslash\{k\}}\frac{x-\beta_{j}}{\beta_{k}-\beta_{j}}, \label{FPMM:LC:encodingB}\IEEEeqnarraynumspace
\end{IEEEeqnarray}
where $\mathbf{Z}_{R+1}^{\mathbf{A}},\ldots,\mathbf{Z}_{R+T}^{\mathbf{A}}$ and $\mathbf{Z}_{R+1}^{\mathbf{B}},\ldots,\mathbf{Z}_{R+T}^{\mathbf{B}}$ are arbitrary matrices over $\mathbb{F}_q$ with the same dimensions as $\mathbf{A}_{r}^{(\theta_1)}$ and $\mathbf{B}_{r}^{(\theta_2)}$ respectively, and will be explained later.
We know from Section \ref{SMM:LC} that, one can recover the desired computation $\mathbf{C}^{(\theta_1,\theta_2)}=\mathbf{A}^{(\theta_1)}\mathbf{B}^{(\theta_2)}$ by interpolating the product polynomial $g(x)=f_{\mathbf{A}}(x)\cdot h_{\mathbf{B}}(x)$.

To keep the indices $\theta_1$ and $\theta_2$ private, the master constructs the query polynomials $\rho_r^{(u)}(x)$ and $q_r^{(v)}(x)$ by exploiting the structure of the encoding functions in \eqref{FPMM:LC:encodingA} and \eqref{FPMM:LC:encodingB}, respectively, for all $r\in[R],u\in[U],v\in[V]$, as follows.
\begin{IEEEeqnarray}{l}
\rho_r^{(u)}(x)=\sum\limits_{k=R+1}^{R+T}\tilde{z}_{r,k}^{(u)}\cdot\prod_{j\in[R+T]\backslash\{k\}}\frac{x-\beta_{j}}{\beta_{k}-\beta_{j}}\notag\\
\quad\quad\quad\quad\quad\quad\quad+\left\{
\begin{array}{@{}l@{\;\;}l}
\prod\limits_{j\in[R+T]\backslash\{r\}}\frac{x-\beta_{j}}{\beta_{r}-\beta_{j}},
&\mathrm{if}\,\, u=\theta_1 \\
0, &\mathrm{if}\,\, u\neq\theta_1
\end{array}
\right., \label{FPMM:LC:queryA}\IEEEeqnarraynumspace\\
q_r^{(v)}(x)=\sum\limits_{k=R+1}^{R+T}z_{r,k}^{(v)}\cdot\prod_{j\in[R+T]\backslash\{k\}}\frac{x-\beta_{j}}{\beta_{k}-\beta_{j}}\notag\\
\quad\quad\quad\quad\quad\quad\quad+\left\{
\begin{array}{@{}l@{\;\;}l}
\prod\limits_{j\in[R+T]\backslash\{r\}}\frac{x-\beta_{j}}{\beta_{r}-\beta_{j}},
&\mathrm{if}\,\, v=\theta_2 \\
0, &\mathrm{if}\,\, v\neq\theta_2
\end{array}
\right.,  \label{FPMM:LC:queryB}\IEEEeqnarraynumspace
\end{IEEEeqnarray}
where $\tilde{z}_{r,k}^{(u)}$ and $z_{r,k}^{(v)}$ are independently and uniformly random noises from $\mathbb{F}_q$.

Then the master evaluates the query polynomials and sends them to worker $i$, i.e.,
\begin{IEEEeqnarray}{rCl}
\mathcal{Q}_{i}^{(\theta_1)}&=&\{\rho_r^{(u)}(\alpha_i):r\in[R],u\in[U]\},\label{FPMM:query:8521}\\
\mathcal{Q}_{i}^{(\theta_2)}&=&\{q_r^{(v)}(\alpha_i):r\in[R],v\in[V]\}.\label{FPMM:query:85211}
\end{IEEEeqnarray}

According to the received queries, worker $i$ encodes its matrices $\mathbf{A}^{([U])}$ and $\mathbf{B}^{([V])}$ into
\begin{IEEEeqnarray}{rCl}
\widetilde{\mathbf{A}}_i&=&\sum\limits_{u=1}^{U}\sum\limits_{r=1}^{R}\mathbf{A}_{r}^{(u)}\cdot \rho_r^{(u)}(\alpha_i),\label{FPMM:LC:encodingU}\\
\widetilde{\mathbf{B}}_i&=&\sum\limits_{v=1}^{V}\sum\limits_{r=1}^{R}\mathbf{B}_{r}^{(v)}\cdot q_r^{(v)}(\alpha_i).\label{FPMM:LC:encodingV}
\end{IEEEeqnarray}

By \eqref{FPMM:LC:encodingA}--\eqref{FPMM:LC:queryB}, we have
\begin{IEEEeqnarray*}{l}
\sum\limits_{u=1}^{U}\!\sum\limits_{r=1}^{R}\mathbf{A}_{r}^{(u)}\!\cdot\! \rho_r^{(u)}(x)\!=\!\sum\limits_{r=1}^{R}\mathbf{A}_{r}^{(\theta_1)}\!\cdot\!\prod_{j\in[R+T]\backslash\{r\}}\frac{x-\beta_{j}}{\beta_{r}-\beta_{j}}\notag\\
\quad\quad\quad\quad\quad\quad\!+\!\sum\limits_{k=R+1}^{R+T}\mathbf{Z}_{k}^{\mathbf{A}}\!\cdot\!\prod_{j\in[R+T]\backslash\{k\}}\frac{x-\beta_{j}}{\beta_{k}-\beta_{j}}=f_{\mathbf{A}}(x),\\
\sum\limits_{v=1}^{V}\!\sum\limits_{r=1}^{R}\mathbf{B}_{r}^{(v)}\!\cdot\! q_r^{(v)}(x)\!=\!\sum\limits_{r=1}^{R}\mathbf{B}_{r}^{(\theta_2)}\!\cdot\!\prod_{j\in[R+T]\backslash\{r\}}\frac{x-\beta_{j}}{\beta_{r}-\beta_{j}}\notag\\
\quad\quad\quad\quad\quad\quad\!+\!\sum\limits_{k=R+1}^{R+T}\mathbf{Z}_{k}^{\mathbf{B}}\!\cdot\!\prod_{j\in[R+T]\backslash\{k\}}\frac{x-\beta_{j}}{\beta_{k}-\beta_{j}}=h_{\mathbf{B}}(x),
\end{IEEEeqnarray*}
where we set
\begin{IEEEeqnarray}{rCl}
\mathbf{Z}_{k}^{\mathbf{A}}=\sum\limits_{u=1}^{U}\sum\limits_{r=1}^{R}\mathbf{A}_{r}^{(u)}\!\cdot\! \tilde{z}_{r,k}^{(u)},\quad\forall\, k\in[R+1:R+T], \notag\\
\mathbf{Z}_{k}^{\mathbf{B}}=\sum\limits_{v=1}^{V}\sum\limits_{r=1}^{R}\mathbf{B}_{r}^{(v)}\!\cdot\! z_{r,k}^{(v)},\quad\forall\, k\in[R+1:R+T],\notag
\end{IEEEeqnarray}
which are constant terms independently of workers.
Then, worker $i$ computes  $\mathbf{Y}_i^{(\theta_1,\theta_2)}=\widetilde{\mathbf{A}}_i\widetilde{\mathbf{B}}_i$, which is the evaluate of the product polynomial $g(x)=f_{\mathbf{A}}(x)\cdot h_{\mathbf{B}}(x)$ at point $x=\alpha_i$.
Thus, the master can interpolate $g(x)$ from the responses of any $K=2R+2T-1$ workers, and then obtains the desired product $\mathbf{C}^{(\theta_1,\theta_2)}=\mathbf{A}^{(\theta_1)}\mathbf{B}^{(\theta_2)}$. 

\subsection{Privacy, Communication Cost and Computation Complexity for FPMM}\label{proof:strategy:2}
For the two FPMM strategies above, their privacy, communication cost, computation complexities and comparisons follow from the similar discussion to the PSMM strategies in Section \ref{proof:strategy}. We briefly outline as follows.
\subsubsection{Privacy}\label{FPMM:Privacy22}

In the two FPMM strategies, the queries $\mathcal{Q}_{i}^{(\theta_1)}$ and $\mathcal{Q}_{i}^{(\theta_2)}$ sent to workers are generated by evaluating the query polynomials $\{\rho_{k,\ell}^{(u)}(x)\}_{k\in[m],\ell\in[p],u\in[U]}$ and $\{q_{\ell,j}^{(v)}(x)\}_{\ell\in[p],j\in[n],v\in[V]}$ (or $\{\rho_r^{(u)}(\alpha_i)\}_{r\in[R],u\in[U]}$ and $\{q_r^{(v)}(\alpha_i)\}_{r\in[R],v\in[V]}$) at distinct points, where each of query polynomials is constructed by employing $T$ independent and uniform random noises to mask interested index, which ensures the privacy of the queries sent to any $T$ workers. Their formal proofs are given in Appendix-B.

\subsubsection{Communication Cost}
In the two strategies, the master downloads a matrix with the same dimension of $\frac{\lambda}{m}\times\frac{\gamma}{n}$ from each of responsive workers.
Thus, the two strategies achieve the download cost  $P_d=\frac{K\times\frac{\lambda}{m}\times\frac{\gamma}{n}}{\lambda\times\gamma}=\frac{K}{mn}$, where $K=\max\{a_{k,\ell},c_t:k\in[m],\ell\in[p],t\in[T]\}+\max\{b_{\ell,j},d_t:\ell\in[p],j\in[n],t\in[T]\}+1$ and $K=2R+2T-1$ for the FPMM strategies based on polynomial codes and Lagrange codes, respectively.

\subsubsection{Computation Complexity}
In the FPMM strategy based on polynomial codes, the complexity at each worker includes encoding the matrices $\mathbf{A}^{([U])}$ by taking a linear combination of $Ump$ sub-matrices with dimension $\frac{\lambda\omega}{mp}$ \eqref{FPMM:answerA}, encoding the matrices $\mathbf{B}^{([V])}$ by taking a linear combination of $Vpn$ sub-matrices with dimension $\frac{\omega\gamma}{np}$ \eqref{FPMM:answerB}, and multiplying the two coded sub-matrices with sizes $\frac{\lambda}{m}\times\frac{\omega}{p}$ and $\frac{\omega}{p}\times\frac{\gamma}{n}$, which requires a complexity of ${O}(U\lambda\omega+V\omega\gamma+\frac{\lambda\omega\gamma}{mpn})$ at most.
Decoding requires interpolating a $(K-1)$-th degree polynomial for $\frac{\lambda\gamma}{mn}$ times, which achieves the complexity ${{O}}(\frac{\lambda\gamma K(\log K)^2\log\log K}{mn})$, where $K=\max\{a_{k,\ell},c_t:k\in[m],\ell\in[p],t\in[T]\}+\max\{b_{\ell,j},d_t:\ell\in[p],j\in[n],t\in[T]\}+1$.

In the FPMM strategy based on Lagrange codes, the complexity at each worker consists of generating the two bathes of sub-matrices $(\mathbf{A}_{1}^{(u)},\ldots,\mathbf{A}_{R}^{(u)}),u\in[U]$ and $(\mathbf{B}_{R}^{(v)},\ldots,\mathbf{B}_{R}^{(v)}),v\in[V]$ by \eqref{FPMM:LC:batchA} and \eqref{FPMM:LC:batchB}, encoding the two batches of sub-matrices into $\widetilde{\mathbf{A}}_i$ and $\widetilde{\mathbf{B}}_i$ by \eqref{FPMM:LC:encodingU}-\eqref{FPMM:LC:encodingV}, and multiplying the two coded sub-matrices $\widetilde{\mathbf{A}}_i$ and $\widetilde{\mathbf{B}}_i$, which achieves the complexity ${O}(UR\lambda\omega+VR\omega\gamma+\frac{\lambda\omega\gamma}{mpn})$.
Decoding is identical to the PSMM strategy based on Lagrange codes, and  achieves the complexity $O(\frac{\lambda\gamma K(\log K)^2\log\log K}{mn}+R\lambda\gamma)$, where $K=2R+2T-1$.

\subsubsection{Performance Comparison Between the Proposed FPMM Strategies}\label{FPMM:COM:2}
The performance of the two FPMM strategies are summarized in Table \ref{tab:FPMM}. Following a discussion similar to Section \ref{com:PSMM:2}, it is straightforward to obtain from Table \ref{tab:FPMM} that, the performance of the FPMM strategy based on polynomial codes strictly outperforms the one based on Lagrange codes when $K_P\leq K_L$ for some values of the parameter $T$, and otherwise the one based on Lagrange codes achieves a smaller recovery threshold and download cost, but still with a higher computation complexity at each worker.

\begin{table*}[htbp]
\centering
\caption{Performance of the proposed FPMM strategies.}\label{tab:FPMM}
  \begin{tabular}{|c|c|c|}
  \hline
  & FPMM  based on Polynomial codes & FPMM  based on Lagrange codes  \\ \hline
\multirow{2}{*}{Recovery Threshold} & $K_P=\max\{a_{k,\ell},c_t:k\in[m],\ell\in[p],t\in[T]\}$ & \multirow{2}{*}{$K_L=2R+2T-1$} \\
& $+\max\{b_{\ell,j},d_t:\ell\in[p],j\in[n],t\in[T]\}+1$ &  \\
 \hline
  Download cost $P_d$ & $\frac{K_P}{mn}$ &  $\frac{K_L}{mn}$  \\ \hline
  Worker Computation ${C}_{w}$ & ${O}(U\lambda\omega+V\omega\gamma+\frac{\lambda\omega\gamma}{mpn})$ & ${O}(UR\lambda\omega+VR\omega\gamma+\frac{\lambda\omega\gamma}{mpn})$  \\ \hline
  Decoding Complexity ${C}_d$ & ${{O}}(\frac{\lambda\gamma K_P(\log K_P)^2\log\log K_P}{mn})$ & $O(\frac{\lambda\gamma K_L(\log K_L)^2\log\log K_L}{mn}+R\lambda\gamma)$  \\ \hline
  \end{tabular}
    \begin{tablenotes}
       \footnotesize
       \item[]\quad\quad\quad\quad\quad\;\; Here, $K_P$ can take the value of $\min\{(m+1)(np+T)-1,(n+1)(mp+T)-1,2mpn+2T-1\}$ by Corollary \ref{cor:PSMM-poly}, \\
       \item[]\quad\quad\quad\quad\quad\;\;  and $R$ denotes any upper construction of bilinear complexity.
    \end{tablenotes}
\end{table*}

\section{Comparison with Related Works}\label{Comparison}
The most valuable aspect of this paper is that we propose a novel systematic approach to construct efficient computation strategies for private matrix multiplication problems.
The key idea is to start with an SMM strategy (polynomial codes-based or Lagrange codes-based), and then carefully design queries at the master such that 1) the interested matrix indices are completely hidden from any $T$ colluding workers, and 2) the response computed from the query and the local data at each worker resembles the response computed in the SMM strategy. Strategies constructed following this approach directly inherit the correctness of matrix multiplication from the underlying SMM strategy, and the original problem is essentially reduced to the problem of designing private queries that are compatible with the chosen SMM strategy, which substantially simplifies the design process for private matrix multiplication strategies. To clearly see the innovations of this approach in perspective, let us compare the strategies constructed following this approach with the previous PSMM strategies \cite{PSMM:1,PSMM:2,PSMM:3,PSMM:4,Qian_Yu} and FPMM strategies \cite{FPMM:1,Qian_Yu} that are most relevant to our work.

References \cite{PSMM:1,PSMM:2,PSMM:3} exploit Polynomial codes  (see, e.g., \cite{Polynomial_code,EP_code}) to solve the PSMM problem without colluding workers (i.e., $T=1$). While both the prior works~\cite{PSMM:1,PSMM:2,PSMM:3} and our first proposed PSMM strategy employ Polynomial codes to encode the confidential matrix $\mathbf{A}$, the main difference lies in how the queries are designed, which accordingly leads to different worker responses and decoding operation. In~\cite{PSMM:1,PSMM:2,PSMM:3}, the queries are designed such that 1) all elements in a query sent to any individual worker have identical distribution, so no information about the interested index $\theta$ is leaked; and 2) the query elements corresponding to $\theta$ are pairwise distinct across all workers for completing the desired matrix multiplication, whereas the remaining elements are made identical to align interference from undesired matrices.
More specifically, in~\cite{PSMM:1,PSMM:2,PSMM:3}, the query sent to worker $i$ is constructed as  $\mathcal{Q}_i^{(\theta)}=\{\alpha_1,\ldots,\alpha_{\theta-1},\alpha_{\theta,i},\alpha_{\theta+1},\ldots,\alpha_V\}$ for any $i\in[N]$, where $\{\alpha_1,\ldots,\alpha_{\theta-1},\alpha_{\theta+1},\ldots,\alpha_V\}$ and $\{\alpha_{\theta,i}\}_{i\in[N]}$ are $N+V-1$ pairwise distinct points that are selected uniformly i.i.d. from $\mathbb{F}_q$.
It is not clear how one can generalize this query design to the colluding case considered in this paper, while maintaining the privacy requirement.
In our proposed strategy, inspired by the specific structure of the polynomial codes utilized to encode the confidential matrix $\mathbf{B}$ in the SMM problem, we design polynomially coded queries (see, e.g., \eqref{query:poly}) that facilitate a form of interference alignment, separating the desired and interfering partitioning sub-matrices in the library $\mathcal{L}^{\mathbf{B}}$, such that the local matrix $\widetilde{{\bf B}}_i$ obtained from the library and the received query has identical structure as the encoding of the matrix $\mathbf{B}$ in SMM, for each worker $i$, as shown in \eqref{Tans:encoding function}. Consequently, the proposed PSMM strategies achieve the same recovery thresholds as the SMM strategies.

For the case of non-colluding workers with $T=1$, the best known recovery threshold achieved by PSMM strategies based on polynomial codes is $mpn+mp+n$~\cite{PSMM:3}. The recovery threshold and download cost of our PSMM strategy based on polynomial codes is superior to that when $m>n$ and $p>1$ by Corollary \ref{cor:PSMM-poly}, with the other performance of upload cost and computation complexity being identical.

The problem of PSMM with $T$ colluding workers was first studied in \cite{PSMM:4}, for the extremely special case of $p=n=1$, i.e., only matrix $\mathbf{A}$ is horizontally divided into $m$ equal-size sub-matrices as $\mathbf{A}_1,\mathbf{A}_2,\ldots,\mathbf{A}_m$.
To complete desired computation, the PSMM strategy based on polynomial codes in \cite{PSMM:4} shares $V$ encoding versions $\mathbf{A}_{i}^{(1)},\mathbf{A}_{i}^{(2)},\ldots,\mathbf{A}_{i}^{(V)}$ of matrix $\mathbf{A}$ to each worker $i$ in a security-preserving manner, which are also used as a query to instruct the worker to compute the response $\mathbf{Y}_i^{(\theta)}=\sum_{v=1}^{V}\mathbf{A}_{i}^{(v)}\mathbf{B}^{(v)}$ for the master, where $\mathbf{A}_{i}^{(v)}$ for any $v\in[V]$ is given by \begin{IEEEeqnarray}{rCl}\notag
\mathbf{A}_{i}^{(v)}=\sum\limits_{t=1}^{T}\mathbf{Z}^{(v)}_t\alpha_i^{t-1}+\left\{
\begin{array}{@{}ll}
\sum\limits_{k=1}^{m}\mathbf{A}_{k}\alpha_i^{T+k-1}, &\mathrm{if}\,\, v=\theta\\
0, & \mathrm{if}\,\, v\neq\theta
\end{array}
\right..
\end{IEEEeqnarray}
Here $\alpha_1,\ldots,\alpha_N$ are distinct elements on $\mathbb{F}_q$ and $\{\mathbf{Z}_t^{(v)}\}_{t\in[T],v\in[V]}$ are random noise matrices. However, this approach requires uploading $V$ encoding versions of $\mathbf{A}$ and pairwise multiplying these encoding sub-matrices with all public matrices in the library $\mathcal{L}^{\mathbf{B}}$ for each worker, implying a significantly communication and computation overheads.
In our PSMM strategy based on polynomial code, the master shares only one secure encoding version $\widetilde{\mathbf{A}}_i$ \eqref{encoding:A} of $\mathbf{A}$, and sends a private query to encode the library  $\mathcal{L}^{\mathbf{B}}$ into one encoding sub-matrix $\widetilde{\mathbf{B}}_i$ \eqref{encoding:B}. The response of each worker is completed by computing the product of the two encoding sub-matrices $\mathbf{Y}_i^{(\theta)}=\widetilde{\mathbf{A}}_i\widetilde{\mathbf{B}}_i$ \eqref{PSMM:PC:response}. For this special case of $p=n=1$, compared with the PSMM strategy in~\cite{PSMM:4}, our proposed strategy reduces upload cost, encoding complexity, and worker computation complexity by a factor of $O(V)$, at the expense of increased recovery threshold by a factor of $2$, where $V$ is the number of matrices in the library $\mathcal{L}^{\mathbf{B}}$ and is typically large in current big data era (see detailed comparison in Table \ref{tab:comparison}).

\begin{table}[htbp]
\centering
\caption{Performance comparison for PSMM strategies based on polynomial codes, for the case of $p=n=1$, and $T > 1$ colluding workers.}\label{tab:comparison}
\resizebox{87mm}{13mm}{
  \begin{tabular}{|@{\;}c@{\;}|@{\;}c@{\;}|@{\;}c@{\;}|@{\;}c@{\;}|}
  \hline
  & \; Previous PSMM Strategy \cite{PSMM:4} &  Our PSMM Strategy \\ \hline
Recovery Threshold  & $K=m+T$ & $K'=2m+2T-1$ \\ \hline
Upload Cost &  $\frac{VN}{m}$ & $\frac{N}{m}$ \\ \hline
Download Cost & $\frac{K}{m}$ & $\frac{K'}{m}$ \\ \hline
Encoding Complexity & $O(\frac{V\lambda\omega N(\log N)^2\log\log N}{m})$ & $O(\frac{\lambda\omega N(\log N)^2\log\log N}{m})$ \\ \hline
Worker Computation & ${O}(\frac{V\lambda\omega\gamma}{m})$ & ${O}(V\omega\gamma+\frac{\lambda\omega\gamma}{m})$ \\ \hline
Decoding Complexity & \; ${{O}}(\frac{\lambda\gamma K(\log K)^2\log\log K}{mn})$ & ${{O}}(\frac{\lambda\gamma K'(\log K')^2\log\log K'}{mn})$ \\ \hline
\end{tabular}}
\end{table}

The FPMM problem with $T$-colluding workers was previously investigated in~\cite{FPMM:1} for the special case of $p=1$, and a strategy with recovery threshold $(m+1)(n+T)+T-1$ is constructed based on the idea of Cross Subspace Alignment (CSA) introduced in~\cite{X-security}.
In the strategy proposed in~\cite{FPMM:1}, the queries sent to worker $i$ are constructed as $\mathcal{Q}_{i}^{(\theta_1)}=\{\rho^{(u)}_k(\alpha_i):k\in[m], u\in[U]\}$ and $\mathcal{Q}_{i}^{(\theta_2)}=\{q^{(v)}_{j}(\alpha_i):j\in[n],v\in[V]\}$, where $\rho^{(u)}_k(\alpha_i)$ and $q^{(v)}_{j}(\alpha_i)$ are given by
\begin{IEEEeqnarray*}{rCl}
\rho^{(u)}_k(\alpha_i)&=&\sum\limits_{t=1}^{T}\tilde{z}_{k,t}^{(u)}\cdot \alpha_i^{t-1}+\left\{
\begin{array}{@{}ll}
\alpha_i^{-(n+T)k}, &\mathrm{if}\,\, u=\theta_1\\
0, & \mathrm{if}\,\, u\neq\theta_1
\end{array}
\right., \\
q^{(v)}_{j}(\alpha_i)&=&\sum\limits_{t=1}^{T}z_{j,t}^{(v)}\cdot \alpha_i^{t-1}+\left\{
\begin{array}{@{}ll}
\alpha_i^{-j}, &\mathrm{if}\,\, v=\theta_2\\
0, & \mathrm{if}\,\, v\neq\theta_2
\end{array}
\right.
\end{IEEEeqnarray*}
for some evaluation point $\alpha_i$ and random noises $\tilde{z}_{k,t}^{(u)},z_{j,t}^{(v)}$. In our FPMM strategy based on polynoimial codes, similar to PSMM, with the goal of resembling the structure of encoding functions for the matrices $\mathbf{A}$ and $\mathbf{B}$ in SMM strategies, our FPMM strategies design corresponding polynomially coded queries (see, e.g., \eqref{query:poly2} and \eqref{query:poly2234}) for the libraries $\mathcal{L}^{\mathbf{A}}$ and $\mathcal{L}^{\mathbf{B}}$, respectively. For this special case of $p=1$, our strategy based on polynomial codes strictly outperforms that in~\cite{FPMM:1} in terms of recovery threshold and download cost by Corollary \ref{cor:PSMM-poly}, while the performance with respect to other measures are identical.

Strategies based on Lagrange codes were first developed in~\cite{Qian_Yu} to solve PSMM and FPMM problems without colluding workers. For this special cases of $T=1$, our proposed strategy based on Lagrange codes achieves identical system performance as that in~\cite{Qian_Yu}, for both problems of PSMM and FPMM. The query design in our proposed strategy differs significantly  from the design in~\cite{Qian_Yu}. While the query design in~\cite{Qian_Yu} follows the ideas in \cite{PSMM:1,PSMM:2} for the PSMM problem, and is difficult to generalize to the case of colluding workers, following our systematic approach, we design Lagrange coded queries of the private indices $\theta_1$ and $\theta_2$, such that the computed response at each worker exhibits identical structure of the response computed using the SMM strategy based on Lagrange codes.

\section{Conclusion}\label{conclusion}
In this paper, we focused on designing efficient PSMM and FPMM strategies that minimize the recovery threshold, communication cost and complexity complexity. 
We showed that, the current SMM strategies based on polynomial codes and Lagrange codes can be used to construct PSMM/FPMM strategies with same recovery threshold, by exploiting the structure inspired by the encoding functions of the SMM strategies to create private queries. This establishes a generic connection between PSMM/FPMM and SMM, and provides a novel systematic approach towards designing PSMM and FPMM strategies. The resulting strategies constructed from this approach improve one or more efficiency metrics including recovery threshold, communication cost and computation complexity, compared with the state of the art, achieving a more flexible tradeoff in optimizing system  efficiency.

\begin{appendix}\label{proof:achevable}

In this appendix, we prove the security and/or privacy of the proposed PSMM/FPMM strategies based on polynomial codes and Lagrange codes.

\subsection{Proof of Security and Privacy for PSMM}
We start with proving the security and privacy of the proposed PSMM strategy based on polynomial codes.
\subsubsection*{Security for PSMM Based on Polynomial Codes}Let $\mathcal{T}=\{i_1,\!\ldots\!,i_T\}\!\subseteq\![N]$ be any $T$ indices of the $N$ workers. Then,
\begin{IEEEeqnarray}{rCl}
&&I(\mathbf{A};\mathcal{Q}_{\mathcal{T}}^{(\theta)},\widetilde{\mathbf{A}}_{\mathcal{T}},\mathbf{B}^{([V])},\mathbf{Y}_{\mathcal{T}}^{(\theta)})\notag\\
&=&I(\mathbf{A};\widetilde{\mathbf{A}}_{\mathcal{T}})+
I(\mathbf{A};\mathcal{Q}_{\mathcal{T}}^{(\theta)},\mathbf{B}^{([V])}|\widetilde{\mathbf{A}}_{\mathcal{T}})\notag\\
&&\quad\quad\quad\quad\quad\quad\quad+I(\mathbf{A};\mathbf{Y}_{\mathcal{T}}^{(\theta)}|\widetilde{\mathbf{A}}_{\mathcal{T}},\mathcal{Q}_{\mathcal{T}}^{(\theta)},\mathbf{B}^{([V])})\notag\\
&\overset{(a)}{=}&I(\mathbf{A};\widetilde{\mathbf{A}}_{\mathcal{T}})\label{proof:security:PC}\\
&\overset{(b)}{=}&I(\{\mathbf{A}_{k,\ell}\}_{k\in[m],\ell\in[p]};f_{\mathbf{A}}(\alpha_i),i\in\mathcal{T})\overset{(c)}{=}0,\notag
\end{IEEEeqnarray}
where $(a)$ is because the queries $\mathcal{Q}_{\mathcal{T}}^{(\theta)}$ and the data matrices $\mathbf{B}^{([V])}$ are generated independently of the encoded matrix $\widetilde{\mathbf{A}}_{\mathcal{T}}$ and the matrix $\mathbf{A}$ by \eqref{PSMM:encodingfunction:A}--\eqref{query:PSMM:PC}, and the responses $\mathbf{Y}_{\mathcal{T}}^{(\theta)}$ \eqref{PSMM:PC:response} are the deterministic function of $\widetilde{\mathbf{A}}_{\mathcal{T}},\mathcal{Q}_{\mathcal{T}}^{(\theta)}$ and $\mathbf{B}^{([V])}$ by \eqref{query:PSMM:PC} and \eqref{encoding:B}, such that $0=I(\widetilde{\mathbf{A}}_{\mathcal{T}},\mathbf{A};\mathcal{Q}_{\mathcal{T}}^{(\theta)},\mathbf{B}^{([V])})\geq I(\mathbf{A};\mathcal{Q}_{\mathcal{T}}^{(\theta)},\mathbf{B}^{([V])}|\widetilde{\mathbf{A}}_{\mathcal{T}})\geq 0$ and
$I(\mathbf{A};\mathbf{Y}_{\mathcal{T}}^{(\theta)}|\widetilde{\mathbf{A}}_{\mathcal{T}},\mathcal{Q}_{\mathcal{T}}^{(\theta)},\mathbf{B}^{([V])})=0$;
$(b)$ follows by \eqref{GT1:partion1} and \eqref{encoding:A};
$(c)$ follows from \eqref{PSMM:encodingfunction:A}, \eqref{SMM:C2} and Lemma \ref{security proof}.

The security of our PSMM strategy based on polynomial codes follows from \eqref{psmm:privacy}.

\subsubsection*{Privacy for PSMM Based on Polynomial Codes}
By \eqref{query:poly} and \eqref{query:PSMM:PC}, the query elements $q_{\ell,j}^{(v)}(\alpha_{i_1}),\ldots,q_{\ell,j}^{(v)}(\alpha_{i_T})$ sent to the workers $\mathcal{T}$ are protected by $T$ random noises $z^{(v)}_{\ell,j,1},\ldots,z^{(v)}_{\ell,j,T}$ chosen independently and uniformly from $\mathbb{F}_q$, for any $\ell\!\in\![p],j\!\in\![n]$ and $v\!\in\![V]$, as shown below.
\begin{IEEEeqnarray}{c}
\left[
\begin{array}{@{}c@{}}
  q_{\ell,j}^{(v)}(\alpha_{i_1}) \\
  q_{\ell,j}^{(v)}(\alpha_{i_2}) \\
  \vdots \\
  q_{\ell,j}^{(v)}(\alpha_{i_T})
\end{array}
\right]\!=\!
\underbrace{\left[
\begin{array}{@{}c@{}}
  h_{\ell,j}^{(v)}(\alpha_{i_1}) \\
  h_{\ell,j}^{(v)}(\alpha_{i_2}) \\
  \vdots \\
  h_{\ell,j}^{(v)}(\alpha_{i_T})
\end{array}
\right]}_{=\mathbf{h}_{\ell,j}^{(v)}}+\underbrace{
\left[
\begin{array}{@{}c@{\;}c@{\;}c@{\;}c@{}}
  \alpha_{i_1}^{d_1} & \alpha_{i_1}^{d_2} & \ldots & \alpha_{i_1}^{d_T} \\
  \alpha_{i_2}^{d_1} & \alpha_{i_2}^{d_2} & \ldots & \alpha_{i_2}^{d_T} \\
  \vdots & \vdots & \ddots & \vdots \\
  \alpha_{i_T}^{d_1} & \alpha_{i_T}^{d_2} & \ldots & \alpha_{i_T}^{d_T} \\
\end{array}
\right]}_{=\mathbf{F}_{\mathcal{T}}}
\underbrace{\left[
\begin{array}{@{}c@{}}
  z^{(v)}_{\ell,j,1} \\
  z^{(v)}_{\ell,j,2} \\
  \vdots \\
  z^{(v)}_{\ell,j,T}
\end{array}
\right]}_{=\mathbf{z}_{\ell,j}^{(v)}},\notag
\end{IEEEeqnarray}
where
\begin{IEEEeqnarray}{rCll}\label{query:poly123}
h_{\ell,j}^{(v)}(x)&=&\left\{
\begin{array}{@{}ll}
x^{b_{\ell,j}}, &\mathrm{if}\,\, v=\theta\\
0, & \mathrm{if}\,\, v\neq\theta
\end{array}
\right..
\end{IEEEeqnarray}

Recall from \eqref{SMM:C2} that $\mathbf{F}_{\mathcal{T}}$ is invertible, whose inverse matrix is denoted by $(\mathbf{F}_{\mathcal{T}})^{-1}$. Then,
\begin{IEEEeqnarray}{rCl}
I(\theta;\mathcal{Q}_{\mathcal{T}}^{(\theta)})&\!\overset{(a)}{=}\!&I(\theta;\{q_{\ell,j}^{(v)}(\alpha_i):i\in\mathcal{T}\}_{\ell\in[p],j\in[n],v\in[V]})\label{PSMM:priv:proof:111}\IEEEeqnarraynumspace\\
&\!=\!&I(\theta;\{\mathbf{h}_{\ell,j}^{(v)}+\mathbf{F}_{\mathcal{T}}\cdot\mathbf{z}_{\ell,j}^{(v)}\}_{\ell\in[p],j\in[n],v\in[V]})\IEEEeqnarraynumspace\\
&\!=\!&I(\theta;\{(\mathbf{F}_{\mathcal{T}})^{-1}\cdot\mathbf{h}_{\ell,j}^{(v)}+\mathbf{z}_{\ell,j}^{(v)}\}_{\ell\in[p],j\in[n],v\in[V]})\IEEEeqnarraynumspace\\
&\!=\!&H(\{(\mathbf{F}_{\mathcal{T}})^{-1}\cdot\mathbf{h}_{\ell,j}^{(v)}+\mathbf{z}_{\ell,j}^{(v)}\}_{\ell\in[p],j\in[n],v\in[V]})\IEEEeqnarraynumspace\\
&&-H(\{(\mathbf{F}_{\mathcal{T}})^{-1}\!\cdot\!\mathbf{h}_{\ell,j}^{(v)}\!+\!\mathbf{z}_{\ell,j}^{(v)}\}_{\ell\in[p],j\in[n],v\in[V]}|\theta)\IEEEeqnarraynumspace\\
&\!\overset{(b)}{=}\!&H(\{(\mathbf{F}_{\mathcal{T}})^{-1}\cdot\mathbf{h}_{\ell,j}^{(v)}+\mathbf{z}_{\ell,j}^{(v)}\}_{\ell\in[p],j\in[n],v\in[V]})\notag\\
&&-H(\{\mathbf{z}_{\ell,j}^{(v)}\}_{\ell\in[p],j\in[n],v\in[V]})\IEEEeqnarraynumspace\\
&\!\overset{(c)}{=}\!&0,\label{PSMM:priv:proof}\IEEEeqnarraynumspace
\end{IEEEeqnarray}
where $(a)$ follows by \eqref{query:PSMM:PC};
$(b)$ holds because $(\mathbf{F}_{\mathcal{T}})^{-1}\cdot\mathbf{h}_{\ell,j}^{(v)}$ are constant numbers by \eqref{query:poly123} when $\theta$ is given, and $\mathbf{z}_{\ell,j}^{(v)}$ are generated independently of $\theta$, for all $\ell\in[p],j\in[n],v\in[V]$, thus $H(\{(\mathbf{F}_{\mathcal{T}})^{-1}\cdot\mathbf{h}_{\ell,j}^{(v)}+\mathbf{z}_{\ell,j}^{(v)}\}_{\ell\in[p],j\in[n],v\in[V]}|\theta)=H(\{\mathbf{z}_{\ell,j}^{(v)}\}_{\ell\in[p],j\in[n],v\in[V]}|\theta)=H(\{\mathbf{z}_{\ell,j}^{(v)}\}_{\ell\in[p],j\in[n],v\in[V]})$;
$(c)$ is due to the fact that all the noises in $\{\mathbf{z}_{\ell,j}^{(v)}\}_{\ell\in[p],j\in[n],v\in[V]}$ are i.i.d. uniformly distributed  on $\mathbb{F}_q$, and are generated independently of $\{(\mathbf{F}_{\mathcal{T}})^{-1}\cdot\mathbf{h}_{\ell,j}^{(v)}\}_{\ell\in[p],j\in[n],v\in[V]}$, such that $\{(\mathbf{F}_{\mathcal{T}})^{-1}\cdot\mathbf{h}_{\ell,j}^{(v)}+\mathbf{z}_{\ell,j}^{(v)}\}_{\ell\in[p],j\in[n],v\in[V]}$ and $\{\mathbf{z}_{\ell,j}^{(v)}\}_{\ell\in[p],j\in[n],v\in[V]}$ are identically and uniformly distributed over $\mathbb{F}_q^{TVpn}$, i.e.,
$H(\{(\mathbf{F}_{\mathcal{T}})^{-1}\cdot\mathbf{h}_{\ell,j}^{(v)}+\mathbf{z}_{\ell,j}^{(v)}\}_{\ell\in[p],j\in[n],v\in[V]})=
H(\{\mathbf{z}_{\ell,j}^{(v)}\}_{\ell\in[p],j\in[n],v\in[V]})=TVpn$.

Further, we have
\begin{IEEEeqnarray}{rCl}
&&I(\theta;\mathcal{Q}_{\mathcal{T}}^{(\theta)},\widetilde{\mathbf{A}}_{\mathcal{T}},\mathbf{B}^{([V])},\mathbf{Y}_{\mathcal{T}}^{(\theta)}) \label{PSMM:SecurityProof1}\\
&=&I(\theta;\mathcal{Q}_{\mathcal{T}}^{(\theta)})+I(\theta;\widetilde{\mathbf{A}}_{\mathcal{T}},\mathbf{B}^{([V])}|\mathcal{Q}_{\mathcal{T}}^{(\theta)})\notag\\
&&\quad\quad\quad\quad\quad\quad\quad+I(\theta;\mathbf{Y}_{\mathcal{T}}^{(\theta)}|\mathcal{Q}_{\mathcal{T}}^{(\theta)},\widetilde{\mathbf{A}}_{\mathcal{T}},\mathbf{B}^{([V])})\IEEEeqnarraynumspace\\
&\overset{(a)}{=}&I(\theta;\mathcal{Q}_{\mathcal{T}}^{(\theta)})\\
&\overset{(b)}{=}&0, \label{PSMM:SecurityProof2}
\end{IEEEeqnarray}
where $(a)$ is similar to \eqref{proof:security:PC} and $(b)$ follows by \eqref{PSMM:priv:proof}.

Thus, the privacy of our PSMM strategy based on polynomial codes follows by \eqref{PSMM:privacy1}.

We next turn to prove the security and privacy of the PSMM strategy based on Lagrange codes. Before that,  a useful lemma is provided.

\begin{Lemma}[Generalized Cauchy Matrix \cite{Lin}]\label{g-cauchy matrix}
Let $\alpha_1,\ldots,\alpha_T$ and $\beta_1,\ldots,\beta_T$ be pairwise distinct elements from $\mathbb{F}_q$, and $v_1,\ldots,v_T$ be $T$ nonzero elements from $\mathbb{F}_q$. Denote by $f_k(x)$ a polynomial of degree $T-1$
\begin{IEEEeqnarray}{c}\notag
f_k(x)=\prod\limits_{j\in[T]\backslash\{k\}}\frac{x-\beta_j}{\beta_k-\beta_j},\quad\forall\,k\in[T].
\end{IEEEeqnarray}
Then the following generalized Cauchy matrix is invertible over $\mathbb{F}_q$.
\begin{IEEEeqnarray}{c}\notag
\left[
  \begin{array}{@{}cccc@{}}
    v_1 f_{1}(\alpha_1) & v_2 f_{2}(\alpha_1)& \ldots & v_T f_{T}(\alpha_1)  \\
    v_1 f_{1}(\alpha_2) & v_2 f_{2}(\alpha_2)& \ldots & v_T f_{T}(\alpha_2)  \\
    \vdots & \vdots & \ddots & \vdots \\
    v_1 f_{1}(\alpha_T) & v_2 f_{2}(\alpha_T)& \ldots & v_T f_{T}(\alpha_T)  \\
\end{array}
\right]_{T\times T}.
\end{IEEEeqnarray}
\end{Lemma}

\subsubsection*{Security for PSMM Based on Lagrange Codes}For any subset $\mathcal{T}=\{i_1,i_2,\ldots,i_T\}\subseteq[N]$ of size $T$,
\begin{IEEEeqnarray}{c}\notag
I(\mathbf{A};\widetilde{\mathbf{A}}_{\mathcal{T}})
\overset{(a)}{=}I(\{\mathbf{A}_{k,\ell}\}_{k\in[m],\ell\in[p]};f_{\mathbf{A}}(\alpha_i),i\in\mathcal{T})
\overset{(b)}{=}0,
\end{IEEEeqnarray}
where $(a)$ follows from \eqref{GT1:partion1} and \eqref{SPMM:encoding:A}; $(b)$ follows by \eqref{BC:AA}, \eqref{PSMM:PL:encdoingAA} and Lemmas \ref{security proof} and \ref{g-cauchy matrix}.

Similar to \eqref{proof:security:PC}, it is straightforward to prove $I(\mathbf{A};\mathcal{Q}_{\mathcal{T}}^{(\theta)},\widetilde{\mathbf{A}}_{\mathcal{T}},\mathbf{B}^{([V])},\mathbf{Y}_{\mathcal{T}}^{(\theta)})=I(\mathbf{A};\widetilde{\mathbf{A}}_{\mathcal{T}})=0$. Thus, the security of our PSMM strategy based on Lagrange codes follows from \eqref{psmm:privacy}.

\subsubsection*{Privacy for PSMM Based on Lagrange Codes}
By \eqref{PSMM:query} and \eqref{SPMM:encoding:AAAA}, the query elements $q_r^{(v)}(\alpha_{i_1}),$ $\ldots,q_r^{(v)}(\alpha_{i_T})$ sent to the workers $\mathcal{T}$ are protected by the $T$ random noises $z^{(v)}_{r,R+1},\ldots,z^{(v)}_{r,R+T}$, for any $r\in[R],v\in[V]$, as follows.
\begin{IEEEeqnarray}{l}
\big[
\begin{array}{@{}c@{\;}c@{\;}c@{}}
  q_r^{(v)}(\alpha_{i_1}) & \ldots & q_r^{(v)}(\alpha_{i_T}) 
\end{array}
\big]^{\mathrm{T}}\!=\!
\big[
\begin{array}{@{}c@{\;}c@{\;}c@{}}
  h_{r}^{(v)}(\alpha_{i_1}) & \ldots &  h_{r}^{(v)}(\alpha_{i_T})
\end{array}
\big]^{\mathrm{T}}\notag\\
\quad+\underbrace{
\left[
\begin{array}{@{}c@{\;\;}c@{\;\;}c@{}}
    v_1(\alpha_{i_1}) f_{1}(\alpha_{i_1}) & \ldots & v_T(\alpha_{i_1}) f_{T}(\alpha_{i_1})  \\
  \vdots & \ddots & \vdots \\
    v_1(\alpha_{i_T}) f_{1}(\alpha_{i_T})& \ldots & v_T(\alpha_{i_T}) f_{T}(\alpha_{i_T})  \\
\end{array}
\right]}_{=\mathbf{F}_{\mathcal{T}}'}
\left[
\begin{array}{@{}c@{}}
  z^{(v)}_{r,R+1} \\
  \vdots \\
  z^{(v)}_{r,R+T}
\end{array}
\right],\notag
\end{IEEEeqnarray}
where
\begin{IEEEeqnarray}{rCll}\notag
h_{r}^{(v)}(x)&=&\left\{
\begin{array}{@{}l@{\;\;}l}
\prod\limits_{j\in[R+T]\backslash\{r\}}\frac{x-\beta_{j}}{\beta_{r}-\beta_{j}},
&\mathrm{if}\,\, v=\theta \\
0, &\mathrm{if}\,\, v\neq\theta
\end{array}
\right.,
\end{IEEEeqnarray}
and 
\begin{IEEEeqnarray}{rCl}\notag
v_k(x)&=&\prod_{j\in[R]}\frac{x-\beta_{j}}{\beta_{R+k}-\beta_{j}},\quad\forall\, k\in[T],\notag\\
f_k(x)&=&\prod_{j\in[R+1:R+T]\backslash\{R+k\}}\frac{x-\beta_{j}}{\beta_{R+k}-\beta_{j}},\quad\forall\, k\in[T].\notag
\end{IEEEeqnarray}

By Lemma \ref{g-cauchy matrix} again, $\mathbf{F}_{\mathcal{T}}'$ is invertible for any subset $\mathcal{T}$. Thus, similar to \eqref{PSMM:priv:proof:111}--\eqref{PSMM:SecurityProof2}, it is easy to prove $I(\theta;\mathcal{Q}_{\mathcal{T}}^{(\theta)},\!\widetilde{\mathbf{A}}_{\mathcal{T}},\!\mathbf{B}^{([V])},\!\mathbf{Y}_{\mathcal{T}}^{(\theta)}) \!=\!I(\theta;\mathcal{Q}_{\mathcal{T}}^{(\theta)})\!=\!I(\theta;\{q_r^{(v)}(\alpha_i)\!:\!i\!\in\!\mathcal{T}\}_{r\in[R],v\in[V]})\!=\!0$ for the PSMM strategy based on Lagrange codes, i.e., our PSMM strategy based on Lagrange codes is private.

\subsection{Proof of Privacy for FPMM}
We next prove the privacy for our two FPMM strategies based on polynomial codes and Lagrange codes.

For any $\mathcal{T}=\{i_1,i_2,\ldots,i_T\}\subseteq[N]$ of size $|\mathcal{T}|=T$, the FPMM strategy based on polynomial codes satisfies
\begin{IEEEeqnarray*}{l}
\quad I(\theta_1,\theta_2;\mathcal{Q}_{\mathcal{T}}^{(\theta_1)},\mathcal{Q}_{\mathcal{T}}^{(\theta_2)},\mathbf{A}^{([U])},\mathbf{B}^{([V])},\mathbf{Y}_{\mathcal{T}}^{(\theta_1,\theta_2)})\\
=\!\!I(\theta_1\!,\!\theta_2;\!\mathcal{Q}_{\mathcal{T}}^{(\theta_1)}\!,\!\mathcal{Q}_{\mathcal{T}}^{(\theta_2)})\!+\!I(\theta_1,\!\theta_2;\mathbf{A}^{([U])},\!\mathbf{B}^{([V])}|\mathcal{Q}_{\mathcal{T}}^{(\theta_1)},\!\mathcal{Q}_{\mathcal{T}}^{(\theta_2)})\\
\quad\quad+I(\theta_1,\theta_2;\mathbf{Y}_{\mathcal{T}}^{(\theta_1,\theta_2)}|\mathcal{Q}_{\mathcal{T}}^{(\theta_1)},\mathcal{Q}_{\mathcal{T}}^{(\theta_2)},\mathbf{A}^{([U])},\mathbf{B}^{([V])})\\
\overset{(a)}{=}\!\!I(\theta_1,\theta_2;\mathcal{Q}_{\mathcal{T}}^{(\theta_1)},\mathcal{Q}_{\mathcal{T}}^{(\theta_2)})\\
\overset{(b)}{=}\!\!I(\theta_1\!,\!\theta_2;\!\{\!\rho_{k,\ell}^{(u)}\!(\!\alpha_i\!),\!q_{\ell,j}^{(v)}\!(\!\alpha_i\!)\!:\!i\!\in\!\mathcal{T}\}_{k\in[m],\ell\in[p],j\in[n],u\in[U],v\in[V]}\!)\\
\overset{(c)}{=}\!\!0,
\end{IEEEeqnarray*}
where $(a)$ follows by the similar argument to \eqref{proof:security:PC}; $(b)$ is due to \eqref{FPMM:queryA} and \eqref{FPMM:queryB};
$(c)$ holds because the query elements $\{\rho_{k,\ell}^{(u)}(\alpha_i):i\in\mathcal{T}\}$ and $\{q_{\ell,j}^{(v)}(\alpha_i):i\in\mathcal{T}\}$ sent to workers $\mathcal{T}$ are protected by random noises $\{\tilde{z}^{(u)}_{k,\ell,t}\}_{t\in[T]}$ and
$\{z^{(v)}_{\ell,j,t}\}_{t\in[T]}$ by \eqref{query:poly2} and \eqref{query:poly2234} respectively, for all
$k\in[m],\ell\in[p],j\in[n],u\in[U],v\in[V]$,
thus $I(\theta_1,\theta_2;\{\rho_{k,\ell}^{(u)}(\alpha_i),q_{\ell,j}^{(v)}(\alpha_i):i\in\mathcal{T}\}_{k\in[m],\ell\in[p],j\in[n],u\in[U],v\in[V]})=0$ follows similar to \eqref{PSMM:priv:proof:111}--\eqref{PSMM:priv:proof}.

Similarly, by \eqref{FPMM:LC:queryA}--\eqref{FPMM:query:85211} and Lemma \ref{g-cauchy matrix}, it is easy to prove that, the FPMM strategy based on Lagrange codes satisfies $I(\theta_1,\theta_2;\mathcal{Q}_{\mathcal{T}}^{(\theta_1)},\mathcal{Q}_{\mathcal{T}}^{(\theta_2)},\mathbf{A}^{([U])},\mathbf{B}^{([V])},\mathbf{Y}_{\mathcal{T}}^{(\theta_1,\theta_2)})=I(\theta_1,\theta_2;\mathcal{Q}_{\mathcal{T}}^{(\theta_1)},\mathcal{Q}_{\mathcal{T}}^{(\theta_2)})=I(\theta_1,\theta_2;\{\rho_r^{(u)}(\alpha_i),q_r^{(v)}(\alpha_i):i\in\mathcal{T}\}_{r\in[R],u\in[U],v\in[V]})=0$.

So the privacy of our two FPMM strategies follows by \eqref{FPMM:privacy:235}.
\end{appendix}

\bibliographystyle{ieeetr}
\bibliography{reference.bib}



\end{document}